\documentclass[12pt]{article}
\usepackage{amsmath}
\usepackage{graphicx}
\usepackage{enumerate}
\usepackage{natbib}
\usepackage{url} 
\usepackage{lineno}
\usepackage{array}
\usepackage{subfigure}
\usepackage{color}

\newcommand{\blind}{1}
\newtheorem{theorem}{Theorem}

\begin{document}

\def\spacingset#1{\renewcommand{\baselinestretch}%
{#1}\small\normalsize} \spacingset{1}


\if1\blind
{
  \title{\bf Effect of Interim Adaptations in Group Sequential Designs}
  \author{Sergey Tarima\\
    Institute for Health and Society, Medical College of Wisconsin\\
    and \\
    Nancy Flournoy \\
    Department of Statistics, University of Missouri-Columbia
  }
  \maketitle
} 
\fi

\if0\blind
{
  \bigskip
  \bigskip
  \bigskip
  \begin{center}
    {\LARGE\bf Effect of interim adaptations in group sequential designs}
  \end{center}
 \medskip
} \fi

\bigskip
\begin{abstract}
This manuscript investigates unconditional and conditional-on-stopping maximum likelihood estimators (MLEs), information measures  and information loss associated with conditioning in group sequential designs (GSDs).  
The possibility of early stopping brings truncation to the distributional form of  MLEs; sequentially, GSD decisions eliminate some events from the sample space.  Multiple testing induces mixtures on the adapted sample space.  Distributions of   MLEs are mixtures of truncated distributions.   Test statistics that are asymptotically normal  without GSD, have  asymptotic distributions, under GSD, that are non-normal  mixtures of truncated normal distributions under local alternatives; under fixed alternatives, asymptotic distributions of test statistics are degenerate.  Estimation of various statistical quantities such as information, information fractions, and confidence intervals  should account for the effect of planned adaptations. Calculation of adapted information fractions requires substantial computational effort. Therefore,  a new GSD is proposed in which stage-specific sample sizes are fully determined by desired operational characteristics, and calculation of information fractions is not needed.
\end{abstract}

\noindent%
\vfill

\newpage
\spacingset{1.5} 

\section{Introduction} \label{Introduction}

Group sequential designs (GSDs) optimize resource allocation and benefit from the possibility of efficacy- and/or 
futility-driven early stopping. Along with these attractive properties, a few challenges require attention from statisticians. Specifically, many group sequential methods rely on the {\em{the joint canonical distribution assumption}} (Section 3.1, \citet{jennison1999})  and/or {\em{Brownian motion theory}} [Section 2.1, \citet{proschan2006}] to justify the choice of a GSD. These assumptions state that \textit{stage-specific standardized test statistics} follow a multivariate normal distribution, or converge to multivariate normality asymptotically. Multivariate normality, however, does not hold for  \textit{cumulative standardized tests statistics} used in common GSD methods including Pocock, O'Brien \& Fleming, and Haybittle-Peto GSDs, not for finite sample sizes, and not asymptotically against local alternatives. The popular SAS's SEQDESIGN procedure, R's gsDesign package, Cytel's EAST and others overcome this difficulty by using a recursive sub-density formula [\citet{Armitage1969}] for assessing type I and power properties. Nonetheless, the impact of non-normality on other statistical quantities such as Fisher information, the information fraction,  and repeated confidence intervals, is not widely recognized.

Multivariate normality of many commonly used stage-specific test statistics $\left(Z_1,\ldots,Z_K\right)$, and consequently their cumulative versions $\mathbf{Z}=\left(Z_{(1)},\ldots,Z_{(K)}\right)$, holds without sequential adaptations. The use of an early stopping criterion, however, eliminates the possibility of some realizations of $\mathbf{Z}$. On its true \textit{adaptation-rule driven support}, the distribution of $\mathbf{Z}$ differs from normal. Thus,  unadapted distributional assumptions should be considered together with the planned adaptation scheme to identify the adaptation-rule driven support of $\mathbf{Z}$ and  its distribution on this adapted support.

Many indications that joint normality does not hold after interim adaptations have been reported. \citet{Demets1994} pointed out that the distribution of $Z_{(k)}$ is not normal and should be estimated recursively. \citet{jennison1999} plotted the density of a normal test statistic in GSD settings, where discontinuity points clearly showed non-normality. \citet{Li2002} found the joint density of stage 1 and stage 2 standardized test statistics not to be bivariate normal. Local asymptotic non-normality was established following sample size recalculations (SSRs) that depend on an interim observed treatment effect [\citet{Tarima2019}]; and a GSD with a single interim analysis can be viewed as a special case of an SSR. Maximum likelihood estimators (MLEs) converging to random mixtures of normal variables have been found in other adaptive designs [\citet{Ivanova2000}, \citet{Ivanova2001}, \citet{May2010}, \citet{Lane2012}].

In studying estimation, several authors have investigated the effect of GSDs on the sample space, bias, uncertainty measures and the amount of information; see \citet{WHITEHEAD1986}, 
\citet{liu1999}, \citet{liu2006}, \citet{Brannath2006}, \citet{schou2013meta},  \citet{milanzi2015},  \citet{Graf2016}, \citet{shimura2017} and \citet{Marschner2018}.

\citet{WHITEHEAD1986} investigated bias in GSDs and suggested a correction. \citet{liu1999} and \citet{liu2006} recognized change in support in the one-parameter exponential family, and investigated unbiased estimation. The adaptation-driven change in support is critical for  derivation MLEs'  distributions  in Section \ref{normality}. \citet{milanzi2015} developed a likelihood approach that applies when the early stopping rule does not depend on the parameter of interest. \citet{schou2013meta} recognized presence of truncation in the joint distribution of stage-specific test statistics. \citet{Brannath2006} and \citet{Graf2016} investigated bias and MSEs in sample size modification problems.  A comprehensive simulation study comparing various GSD estimators of a parameter of interest is reported in \citet{shimura2017}. \citet{Marschner2018} have found information in conditional and unconditional GSD MLEs and quantified loss associated with conditioning. As it is shown in Section \ref{normality}, unconditional and conditional information measures differ from what they reported in supplementary material to their manuscript, but the information loss is the same.

However, the normality and asymptotic normality assumption continues to be directly used with non-normally distributed statistics. \citet{mehta2007} built repeated confidence intervals. \citet{koopmeiners2012} explored MLEs conditional on stopping but assumed asymptotic normality to evaluate their uncertainty. \citet{martens2018} relied on asymptotic normality for evaluating regression coefficients under the Fine--Gray model in GSD settings. \citet{Asendorf2018} evaluated asymptotic properties with SSR under a fixed alternative  for negative binomial random variables. \citet{gao2013} is a rare exception  in not making a normality assumption; these authors mostly dealt with set operations and probabilities and, using stage-wise ordering of events, they calculated P-values, confidence intervals, and a median unbiased estimate of the parameter of interest.

Asymptotic normality holds when a sequential study is powered against a fixed (not local) alternative. The problem  is that  all consistent tests asymptotically reject the null in favor of \emph{any} fixed alternative hypothesis with $100\%$ statistical power. As the sample size increases, with a proportional increase of all stage-specific sample sizes, the probability to reject the  converges to $1$ at stage 1 and tests statistics degenerate to a point mass, see Section 7.4 in \citet{Fleming1991}. Thus, for large samples at fixed alternatives, the need   for a power concept and the need for an interim analysis disappears.

To mitigate the limited applicability of multivariate normality to GSD trials with possibility of early stopping, many researchers rely on a sequential recursive formula suggested by \citet{Armitage1969} to find the distribution of $Z_{(k)}$ conditional on reaching stage $k$.

In this manuscript, Section \ref{normality} introduces notation for  GSDs with  stopping rules dependent on a parameter of interest and shows through a few examples of two-stage GSDs that conditional and unconditional distributions are truncated or mixtures of truncated distributions. Fisher information in MLEs is also derived in this section. Section \ref{mixtureD} presents a method for designing a sequential experiment controlling a pre-determined power against a sequence of ordered alternatives while controlling type 1 error. This new approach does not use information fraction arguments and the design is fully driven by desired operational characteristics. Section \ref{LSP} derives a local asymptotic distribution of the MLEs, which is a mixture of truncated normal distributions. Section \ref{example} shows an illustrative application of the theory. Finally, Section \ref{Conclusion} concludes this article with a short summary.

\section{Likelihood in Group Sequential Designs} \label{normality}
Consider a random variable $X$ with a p.d.f or a p.m.f. $f_X\left(x|\theta\right)$ and the objective of testing the null hypothesis $H_0:\theta=0$ with $\alpha$-level type~1 error and $1-\beta$ power at an alternative $H_1: \theta=\theta_1$. In GSDs,  it is convenient to  group the random sample in vectors corresponding  interim analyses: $\mathbf{X}_{1}, \ldots, \mathbf{X}_{K}$, where $\mathbf{X}_{k} = (X_{n_{(k-1)}+1},
\ldots,X_{n_{(k)}})$; $n_{(k)}=\sum_{i=1}^kn_i$; $n_i$ is a total number of observations in a stage $i$, $1\le i \le  k\le K$, and $n_{(0)}=0$. Each stage  terminates with an analysis that results in a decision to stop the study or to enroll a new group of patients.  Every GSD stage is assumed to be ``reachable'',  that is, there is a positive probability of reaching each stage. Further, to simplify the material, the term \textit{density} is used to refer to probability measures without formally distinguishing between p.m.f.s and p.d.f.s.

The assumption of a joint canonical distribution states that $\mathbf{Z}$ follows a multivariate normal distribution with a vector of means $\left(\theta\sqrt{{\cal{I}}_1},\ldots,\theta\sqrt{{\cal{I}}_K}\right)$ and a covariance matrix with elements $\sqrt{{\cal{I}}_{k_1}/{\cal{I}}_{k_2}}$, $1 \le k_1\le k_2 \le K$ [Section 3.1 in \citet{jennison1999}]. Normal approximations are useful for many statistical tests. For example, if a researcher is interested in estimating $\theta$ with i.i.d.~observations and the MLE of $\theta$ at the end of stage~$k$  is $\widehat\theta_{k}$, then under common regularity conditions, $\sqrt{n_k}(\widehat\theta_{k}-\theta)/\sigma_{\theta}\to N(0,1)$ and the statistic $Z_k = \sqrt{n_k}\ \widehat\theta_{k}/\sigma_{\theta}\to N(0,1)$ when $\theta = 0$. Then if the sample size  is large enough,   a  normal approximation is reasonable. But when the $Z_k$ is combined with $Z_1,\ldots,Z_{k-1}$ in a pooled test statistic $Z_{(k)}$, the possibility of early stopping breaks the joint canonical assumption and makes $Z_{(k)}$ non-normally distributed. This non-normality is illustrated with the example of  Pocock's two-stage design in Section \ref{Pocock_example}.
Now a few \textit{finite sample} results are presented;  asymptotic results are reported  in Section \ref{LSP}.

\subsection{Conditional and Unconditional Maximum Likelihood Estimation} \label{Section2.3}

In GSDs, $\alpha$-spending functions determine cutoff values $\{c_1,\ldots,c_K\}$ that drive decisions to stop at stage $k$ (for $1\le k < K-1$) or continue  through stage~$K$.  These decisions are defined by the events $\left\{\cap_{j=1}^{k-1}\left\{Z_{(j)}\le c_j\right\}\right\}\cap \left\{Z_{(k)}>c_k\right\}$ and $\cap_{j=1}^{K-1}\left\{Z_{(j)}\le c_k\right\}$, respectively, and they are conveniently summarized by a random variable denoting the stopping stage:
$$D := K \cdot I\left(\cap_{j=1}^{K-1}\left\{Z_{(j)}\le c_j\right\}\right) 
+ \sum_{k=1}^{K-1} k \cdot I\left(\left(\cap_{j=1}^{k-1}\left\{Z_{(j)}\le c_j\right\}\right)\cap \left\{Z_{(k)}>c_k\right\}\right)$$
defined on $1,\ldots,K$.   

The statistic $Z_{(k)}$ is a function of the  observations $\textbf{\textit{X}}_{(k)} = (X_1,\ldots,X_{n_{(k)}})$ that were observed prior to stopping.
$D$ will appear as random index, such as  in $\mathbf{X}_{(D)}$, to underline the fact that the stopping stage is unknown and is described probabilistically though the random variable $D$. 
The fixed index $k$ in $\mathbf{X}_{(k)}$ indicates that the random variable $D$ took value $k$ (the experiment stopped at stage $k$). Thus, $\mathbf{X}_{(k)}$ is the  random variable $\mathbf{X}_{(D)}$ conditioned on $D=k$.
Given $D=k$, the values  $\{x_j:j \ge n_{(k)}\}$ are unobserved and hence do not contribute to the density; indeed, they  do not belong to the adaptation-rule driven sample space, and consequently, they do not belong to a $\sigma$-field defined on the experiment's sample space, measurable space, probability space, and hence, to the statistical experiment as a whole. Some researchers view $\{x_j:j \ge n_{(k)}\}$ as missing data, but in this paper, by analogy with structural zeroes in contingency tables, these values are excluded from the sample space.

\begin{figure}[b!]
\centering
\subfigure[Sketch of support components by  (scaleless) stage-specific test statistics]{\includegraphics[width = 2.5in]{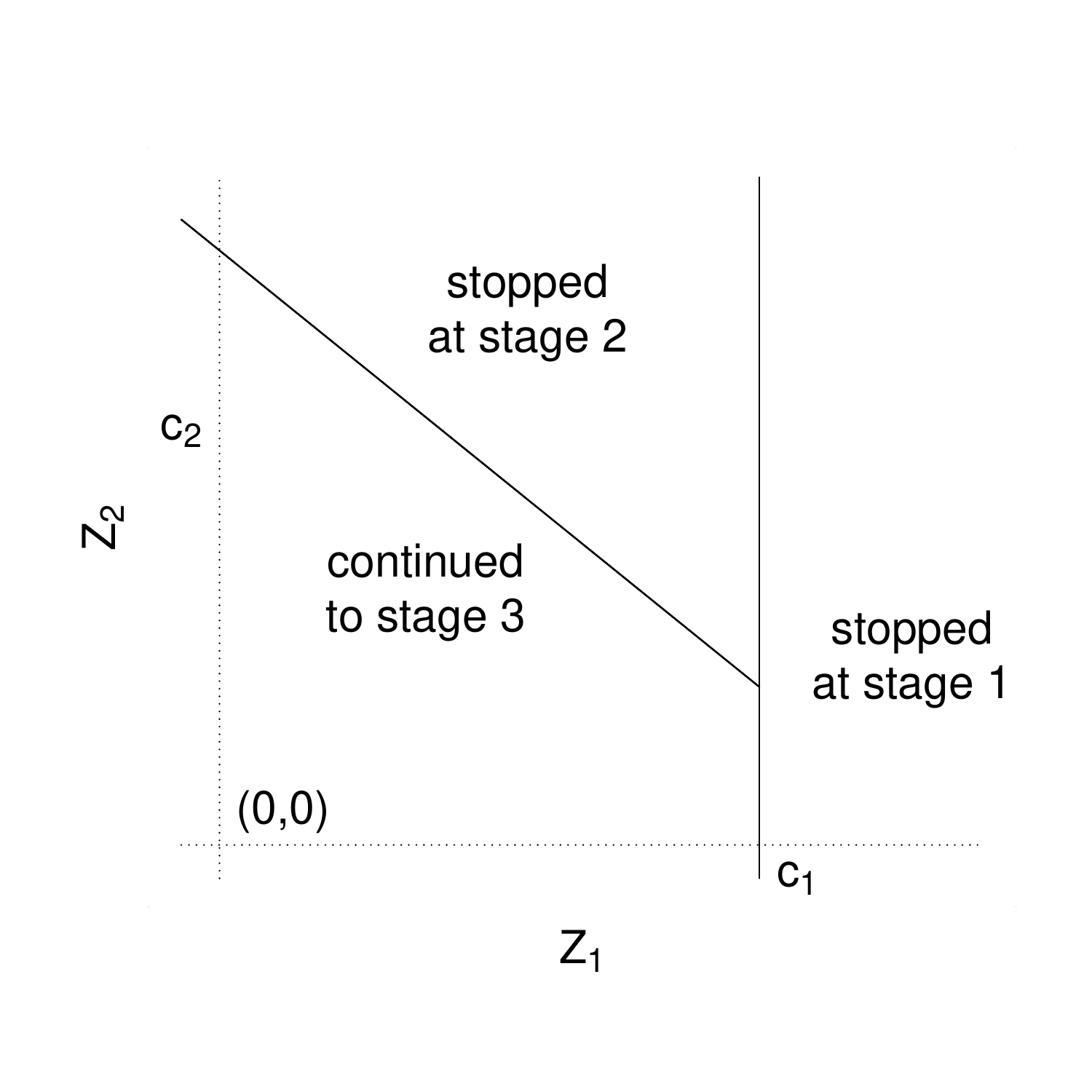}} \quad\quad
\subfigure[Sketch of support components by  (scaleless) cumulative test statistics]{\includegraphics[width = 2.5in]{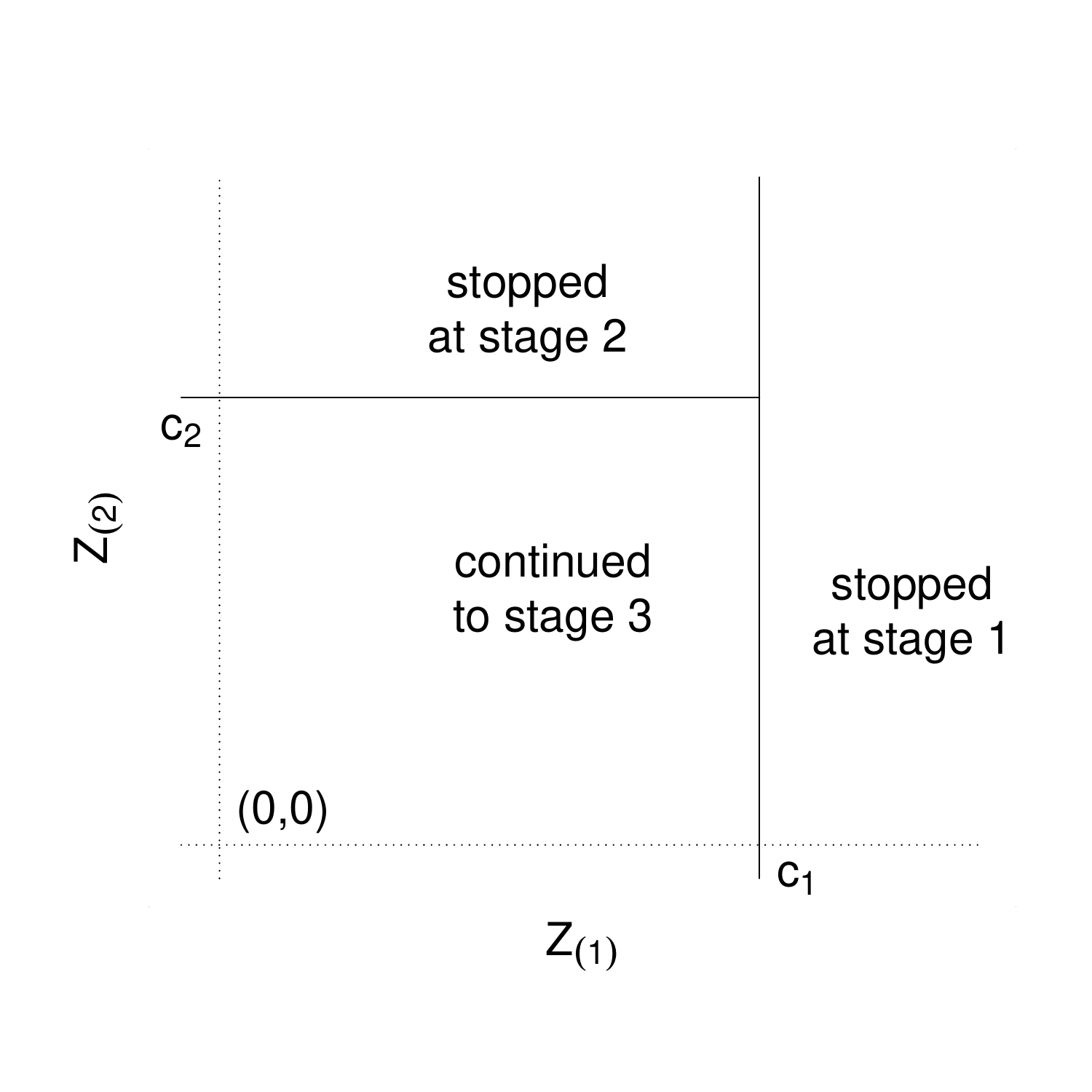}}\\
\caption{\label{support_white}  Support  associated with different stopping decisions when $K=3$;
 $c_k$ is the critical value for stopping at stage $k$.}
\end{figure}

Now the \textit{joint} density for this GSD can be written as
\begin{eqnarray} \label{densjoint}
f_{\mathbf{X}_{(D)}}\left(\textbf{\textit{x}}_{(D)}\vert \theta \right) :=
\prod_{k=1}^{K} [f_{\mathbf{X}_{(k)}}(\textbf{\textit{x}}_{(k)}|\theta)]^{I\left(D=k\right)}
=\prod_{k=1}^{K}f_{\mathbf{X}_{(k)}}^{sub}
,\end{eqnarray} 
where 
$f_{\mathbf{X}_{(k)}}^{sub}:=[f_{\mathbf{X}_{(k)}}(\textbf{\textit{x}}_{(k)}|\theta)]^{I\left(D=k\right)}$ denotes \textit{sub-densities} with support defined by $D=k$, $k=1,\ldots,K$. 
The joint density (\ref{densjoint}) is a mixture with its components defined on non-overlapping regions of the density's support. 
Considering the density \eqref{densjoint} conditional on the observed data, $\left(k,\textbf{\textit{x}}_{(k)}\right)$, the  (unconditional) likelihood is
\begin{eqnarray} \label{LKjoint}
{\mathcal{L}}\left(\theta \vert k,\textbf{\textit{x}}_{(k)} \right) =
f_{\mathbf{X}_{(k)}}^{sub}.\end{eqnarray} 
In contrast, the density of  observations conditional on stopping at stage $k$, i.e., the density of  $ \mathbf{X}_{(k)}=\mathbf{X}_{(D)}\vert \{D=k\}$, 
 is
\begin{align}\label{densK}
f^c_{\mathbf{X}_{(k)}} 
&:= f_{\mathbf{X}_{(D)}} \left(\textbf{\textit{x}}_{(D)}\vert D=k, \theta \right) 
= \frac{I\left(D=k\right)}{\text{Pr}_{\theta}\left(D=k\right)} f_{\mathbf{X}_{(k)}} \left(\textbf{\textit{x}}_{(k)}\vert  \theta \right)
= I\left(D=k\right)\frac{f_{\mathbf{X}_{(k)}}^{sub}}{\text{Pr}_{\theta}\left(D=k\right)}\end{align}
with associated  likelihood 
\begin{eqnarray} \label{LKcon}
{\mathcal{L}}^c\left(\theta \vert k,\textbf{\textit{x}}_{(k)}  \right) =
 I\left(D=k\right)\frac{f_{\mathbf{X}_{(k)}}^{sub}}{\text{Pr}_{\theta}\left(D=k\right)}.\end{eqnarray} 
The indicator function in \eqref{densK} and \eqref{LKcon} emphasizes that support for the random variables is reduced by the conditioning (see Figure~\ref{support_white}).
Note also in Figure~\ref{support_white} that the support conditional on stopping at one stage is disjoint from the support conditional on stopping at another stage.
If the stopping rule is not random and the experiment stops with $n_{(k)}$ observations, $\text{Pr}_{\theta}(D=k)\equiv 1$ and
\begin{eqnarray} \label{Lfix}
{\mathcal{L}}^{fix}\left(\theta \vert k,\textbf{\textit{x}}_{(k)}  \right) =
f_{\mathbf{X}_{(k)}}.\end{eqnarray}
Both ${\cal{L}}$ and ${\cal{L}}^c$ are functions of $\theta$ and the observed data $(k, \textbf{\textit{x}}_{(k)})$, but ${\cal{L}}$ and ${\cal{L}}^c$ are continuous functions  only of $\theta$ and have discontinuities in $(k, \textbf{\textit{x}}_{(k)})$ since the distribution of $\mathbf{X}_{(D)}$ is a mixture. These discontinuities are inherited by unconditional and conditional MLEs
 $\widehat\theta$ and $\widehat\theta^c$, respectively, that are defined below. Adaptation can change distributions of the MLEs and their information measures.

Conditional on $D=k$, MLEs maximizing ${\mathcal{L}}$ and ${\mathcal{L}}^c$, respectively, 
 are
 \begin{align*}
\widehat\theta_{(k)} &= \arg \max_{\theta} f_{\mathbf{X}_{(k)}}^{sub} \quad \textrm{ and } \quad 
\widehat\theta^{c}_{(k)} = \arg \max_{\theta} \left[I\left(D=k\right) \frac{f_{\mathbf{X}_{(k)}}^{sub}}{\text{Pr}_{\theta}\left(D=k\right)}\right].
\end{align*}
The estimator $\widehat\theta^{c}_{(k)}$ was suggested by \citet{koopmeiners2012}.
Overall, the  MLEs derived from \eqref{densjoint} and \eqref{densK}, respectively, can be written as
$$ \widehat\theta = \sum_{k=1}^K I(D=k) \widehat\theta_{(k)}\quad\textrm { and }\quad
\widehat\theta^c = \sum_{k=1}^K I(D=k) \widehat\theta_{(k)}^c.$$

For every observed pair $\left(k,\textbf{\textit{x}}_{(k)}\right)$, $f_{\mathbf{X}_{(k)}}^{sub} =f_{\mathbf{X}_{(k)}}$ for all $\theta$ and, consequently, 
$$\widehat\theta_{(k)}^{fix} := \arg \max_{\theta} f_{\mathbf{X}_{(k)}} = \arg \max_{\theta} f_{\mathbf{X}_{(k)}}^{sub} = \widehat\theta_{(k)}$$ and 
$$ \widehat\theta^{fix}=\sum_{k=1}^K I(D=k) \widehat\theta_{(k)}^{fix} 
= \sum_{k=1}^K I(D=k) \widehat\theta_{(k)}=
\widehat\theta,   $$ 
which means that MLEs stay unchanged for maximization of either ${\mathcal{L}}$ and ${\mathcal{L}}^{fix}$.

\subsection{Conditional and Unconditional Information}

 The superscript \textit{obs} is used to denote the observed  information (and quantities obtained from it); and  the superscript \textit{fix} is used to denote the quantities  derived assuming that study stops at a pre-determined fixed sample size (no adaptation). As previously, the superscript  \textit{c}  denotes quantities obtained from the conditional likelihood \eqref{LKcon} and   the stopping stage is referenced, when needed, in subscripts.  

With this notation, the \emph{observed information measures} derived from ${\cal{L}}$ \eqref{LKjoint}, ${\cal{L}}^c$\eqref{LKcon}, and ${\cal{L}}^{fix}$\eqref{Lfix}, respectively, are 
\begin{align}\label{obsI}
{\cal{I}}^{obs}_{(k)} 
&=-\frac{\partial^2}{\partial\theta^2} \log  f_{\mathbf{X}_{(k)}}^{sub}(\textbf{\textit{x}}_{(k)}|\theta)
=-\frac{\partial^2}{\partial\theta^2}  I(D=k)\log  f_{\mathbf{X}_{(k)}}(\textbf{\textit{x}}_{(k)}|\theta),
\\
{\cal{I}}^{obs,c}_{(k)}  
&={\cal{I}}^{obs}_{(k)} +\frac{\partial^2}{\partial\theta^2} \log\text{Pr}_{\theta}\left(D=k\right) \quad\textrm{ and }\\
{\cal{I}}^{obs,fix}_{(k)}&=-\frac{\partial^2}{\partial\theta^2} \log  f_{\mathbf{X}_{(k)}}.
\end{align}
Observed information matrices evaluated at MLEs are positive definite.  Therefore, \\$\frac{\partial^2}{\partial\theta^2} \log\text{Pr}_{\theta}\left(D=k\right)\vert_{\theta=\widehat\theta}<0$ and
${\cal{I}}^{obs,c}_{(k)}\vert_{\theta=\widehat\theta} < {\cal{I}}^{obs}_{(k)}\vert_{\theta=\widehat\theta}$.

\emph{Stopped expected information measures} are
\begin{align*}
 {\cal{I}}_{(k)}&:= \text{E}_{\textbf{X}_{(k)}}\left[{\cal{I}}^{obs}_{(k)}\right]
=\int \cdots \int {\cal{I}}^{obs}_{(k)}
\frac{f_{\mathbf{X}_{(k)}}^{sub}(\textbf{\textit{x}}_{(k)}|\theta)}{\text{Pr}_{\theta}(D=k)}\prod_{i=1}^{n_{(k)}} dx_i \quad\textrm{ and }\\
{\cal{I}}^c_{(k)}&:= \text{E}_{\textbf{X}_{(k)}}\left[{\cal{I}}^{obs,c}_{(k)}\right] =\int \cdots \int \left[ {\cal{I}}^{obs}_{(k)} +\frac{\partial^2}{\partial\theta^2} \log\text{Pr}_{\theta}\left(D=k\right)\right]
 \frac{f_{\mathbf{X}_{(k)}}^{sub}(\textbf{\textit{x}}_{(k)}|\theta)}{\text{Pr}_{\theta}(D=k)}\prod_{i=1}^{n_{(k)}} dx_i.
\end{align*}
Under $\text{Pr}_{\theta}(D=k)\equiv 1$, 
${\cal{I}}_{(k)}$ and ${\cal{I}}_{(k)}^c$ reduce to
Fisher information from $n_{(k)}$ observations: 
\begin{align}\label{I_(k)2}
 {\cal{I}}_{(k)}^{fix}&=\int \cdots \int \left[-\frac{\partial^2}{\partial\theta^2} \log  f_{\mathbf{X}_{(k)}}(\textbf{\textit{x}}_{(k)}|\theta) \right]
f_{\mathbf{X}_{(k)}}(\textbf{\textit{x}}_{(k)}|\theta)\prod_{i=1}^{n_{(k)}} dx_i.
\end{align}%
Overall, the expected (Fisher) information measures   are
\begin{align}
{\cal{I}} &= \text{E}\left[{\cal{I}}^{obs}_{(D)}\right] 
=
 \text{E}_{D}\left[{\cal{I}}_{(k)}\right] 
=\sum_{k=1}^D \text{Pr}_{\theta}(D=k) \text{E}_{\textbf{X}_{(k)}}\left[{\cal{I}}^{obs}_{(k)}\right] \quad\textrm{ and } \notag\\
{\cal{I}}^c &=\text{E}\left[{\cal{I}}^{obs,c}_{(D)}\right]
=\text{E}_D\left[{\cal{I}}^{c}_{(k)}\right]
= \sum_{k=1}^D \text{Pr}_{\theta}(D=k) \text{E}_{\textbf{X}_{(k)}}\left[{\cal{I}}^{obs,c}_{(k)}\right]\notag \\
&= {\cal{I}} - \text{E}_D\left[-\frac{\partial^2}{\partial\theta^2} \log  \text{Pr}_{\theta}\left(D\right)\right].\label{e6}
\end{align}
Note that the information loss incurred by conditioning is equal to the amount of Fisher information about $\theta$  in the early stopping rule, namely $\text{E}_D\left[-\frac{\partial^2}{\partial\theta^2} \log  \text{Pr}_{\theta}\left(D\right)\right]$. If ${\text{Pr}}_{\theta}\left(D\right)$ does not depend on $\theta$,  ${\cal{I}}={\cal{I}}^{c}$.
If the fact that ${\cal{I}}_{(k)}^{fix}$ assumes $\text{Pr}_{\theta}(D=k)\equiv 1$ is ignored, the use of ${\cal{I}}_{(k)}^{fix}$ with every $D=k$ leads to 
$${\cal{I}}^{fix} := \sum_{k=1}^K \text{Pr}_{\theta}\left(D=k\right)  {\cal{I}}_{(k)}^{fix}.$$ 

Since $f_{\mathbf{X}_{(k)}}^{sub}(\textbf{\textit{x}}_{(k)}|\theta)$ can be positive only when $D=k$, and 
$f_{\mathbf{X}_{(k)}}^{sub}(\textbf{\textit{x}}_{(k)}|\theta) = f_{\mathbf{X}_{(k)}}(\textbf{\textit{x}}_{(k)}|\theta)$ when $D=k$, ${\cal{I}}_{(k)}$ is re-written as
\begin{eqnarray} 
{\cal{I}}_{(k)} 
&=& \underset{I(D=k)}{\int \cdots \int} \left[-\frac{\partial^2}{\partial\theta^2} \log  f_{\mathbf{X}_{(k)}}(\textbf{\textit{x}}_{(k)}|\theta) \right]
\frac{f_{\mathbf{X}_{(k)}}(\textbf{\textit{x}}_{(k)}|\theta)}{\text{Pr}_{\theta}(D=k)}\prod_{i=1}^{n_{(k)}} dx_i\label{e7}, 
\end{eqnarray}
From (\ref{I_(k)2}) and (\ref{e7}), ${\cal{I}}_{(k)} \ne {\cal{I}}_{(k)}^{fix}$ under GSDs. At $\theta=\widehat\theta$, 
$\frac{\partial^2}{\partial\theta^2}\log  f_{\mathbf{X}_{(k)}}\left(\textbf{\textit{x}}_{(k)}|\theta=\widehat\theta\right) \le 0$ and ${\cal{I}}_{(k)} \le {\cal{I}}_{(k)}^{fix}$ and ${\cal{I}} \le {\cal{I}}^{fix}$.

\subsection{Example: Pocock One-Sided Two-Group Sequential Z-test}\label{Pocock_example}

Pocock's design is a simple two-stage study design for testing $H_0: \theta = 0$ versus $H_1: \theta = \theta_1$ with $X_i \sim N(\theta,1)$.  Under $n_1=n_2=100$,$c_1=2.18$ is used to secure an overall type~1 error rate $\alpha=0.025$ with a one sided $z$ test. If $Z_1 \le 2.18$,  $Z_2$ is also observed, where $$Z_k = \frac{1}{\sqrt{n_{k}}}\sum_{i=n_{(k-1)}+1}^{n_{(k)}} X_{i} = \sqrt{n_{k}} \cdot \bar X_k  \overset{d}{=} N(\theta,1).$$ If there is no possibility of interim stopping, Fisher information from both stages combined is ${\cal{I}}_{(2)}^{fix}=200$. This directly follows from the additivity property of Fisher information for independent data.

\begin{figure}[htb!]
\centering
\subfigure[$\widehat\theta_{(1)}$]{\includegraphics[width = 2in,height=.18\textheight]{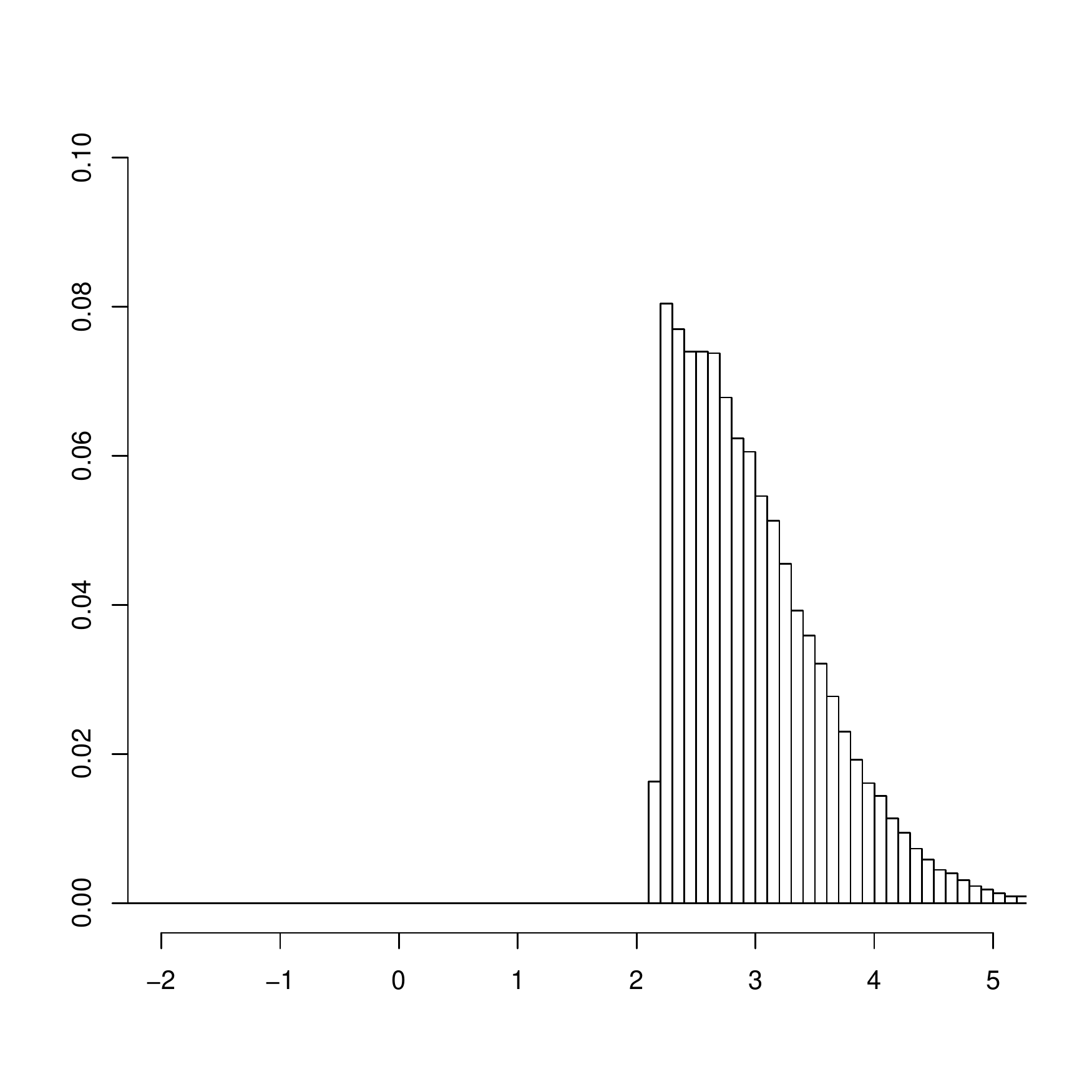}} 
\subfigure[$\widehat\theta_{(2)}$]{\includegraphics[width = 2in,height=.18\textheight]{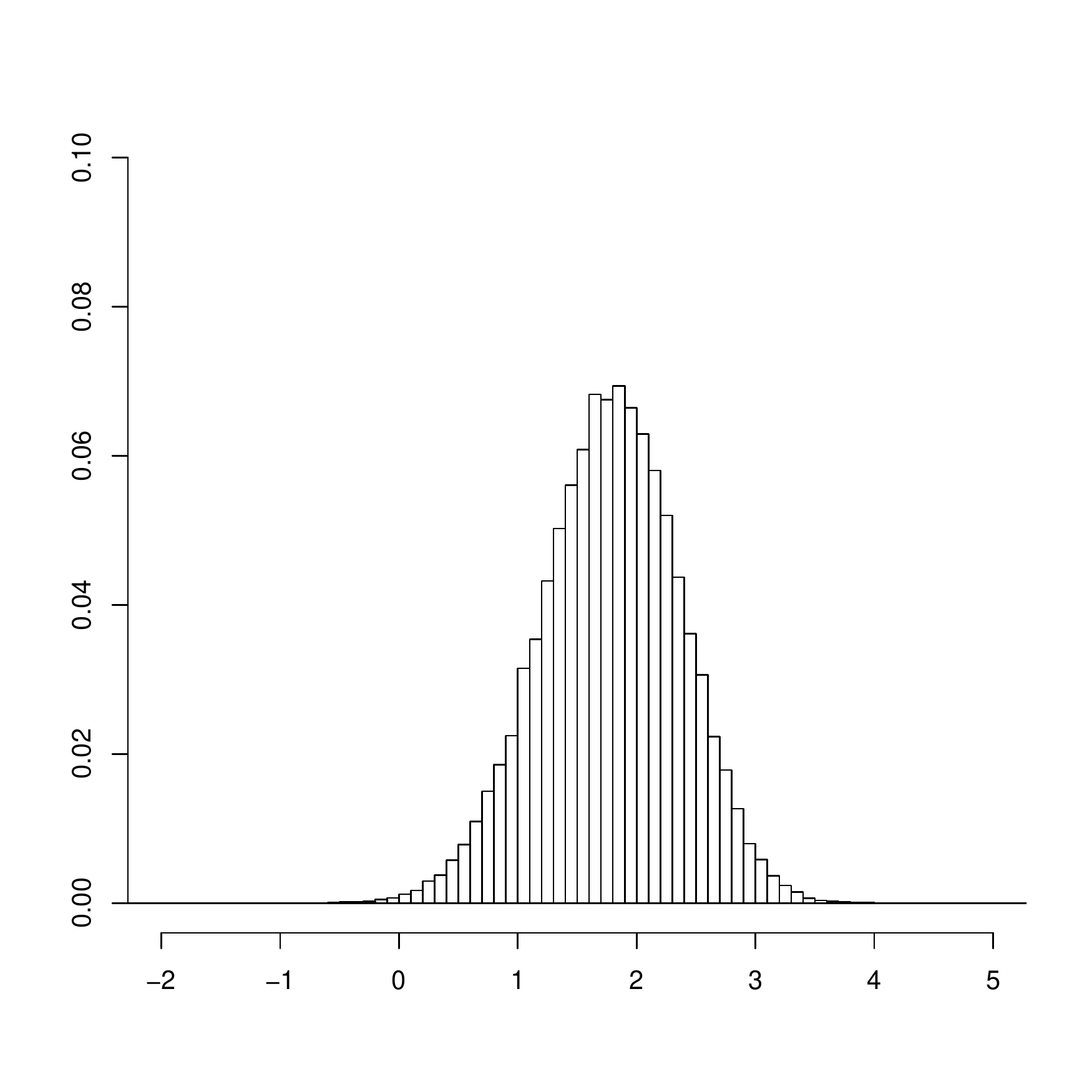}} 
\subfigure[$\widehat\theta$, a mixture of $\widehat\theta_{(1)}$ and $\widehat\theta_{(2)}$]{\includegraphics[width = 2in,height=.18\textheight]{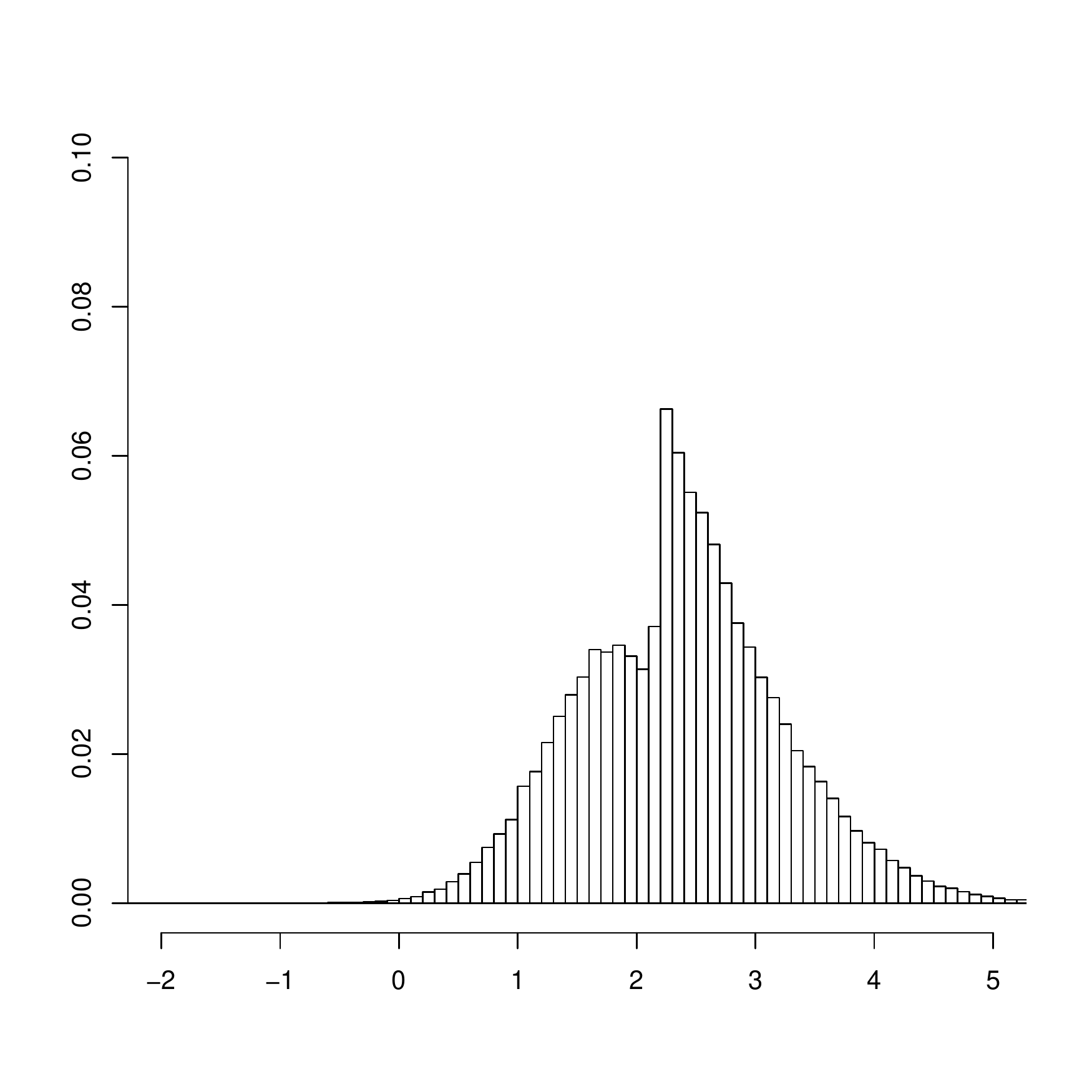}} \\
\subfigure[$\widehat\theta_{(1)}^c$]{\includegraphics[width = 2in,height=.18\textheight]{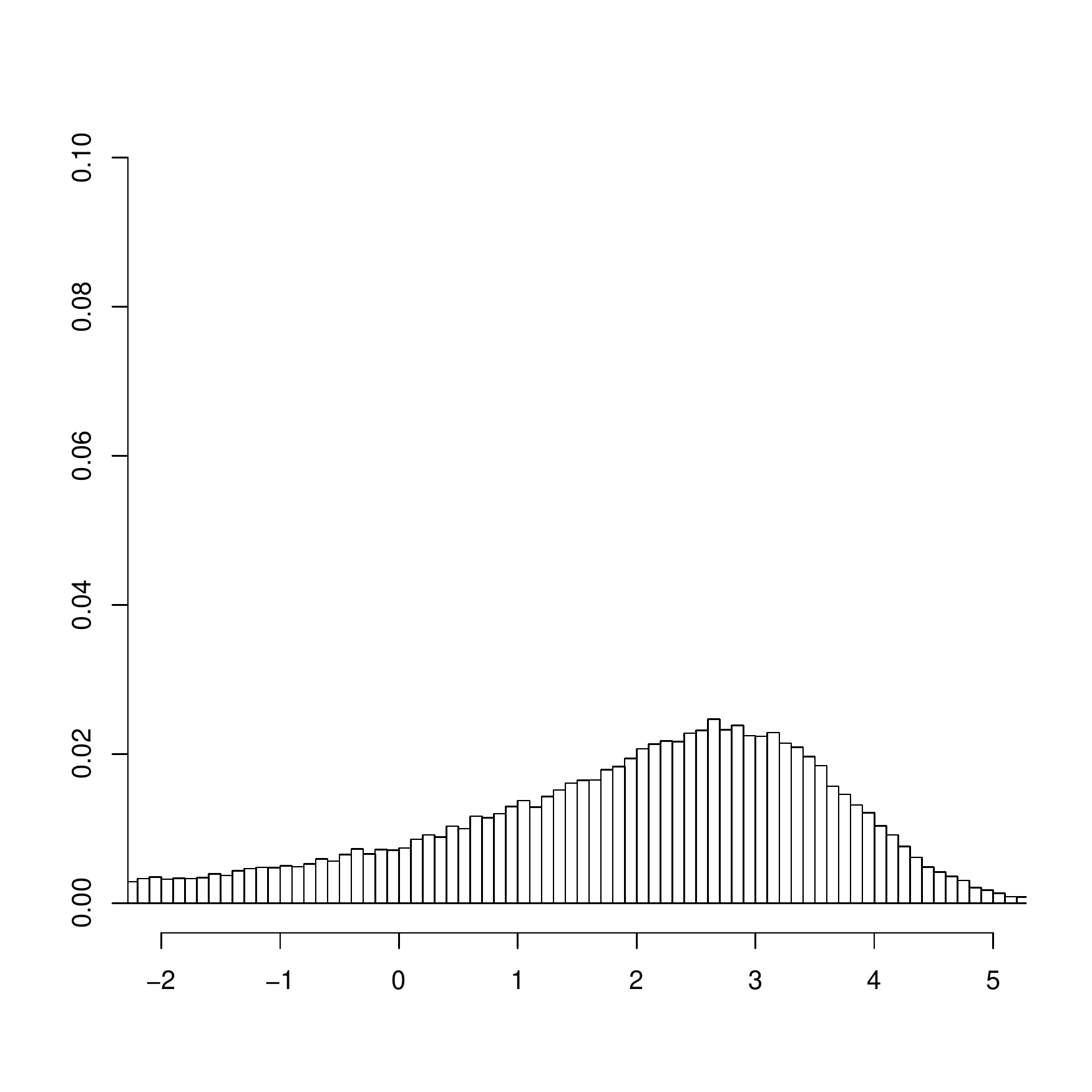}}
\subfigure[$\widehat\theta_{(2)}^c$]{\includegraphics[width = 2in,height=.18\textheight]{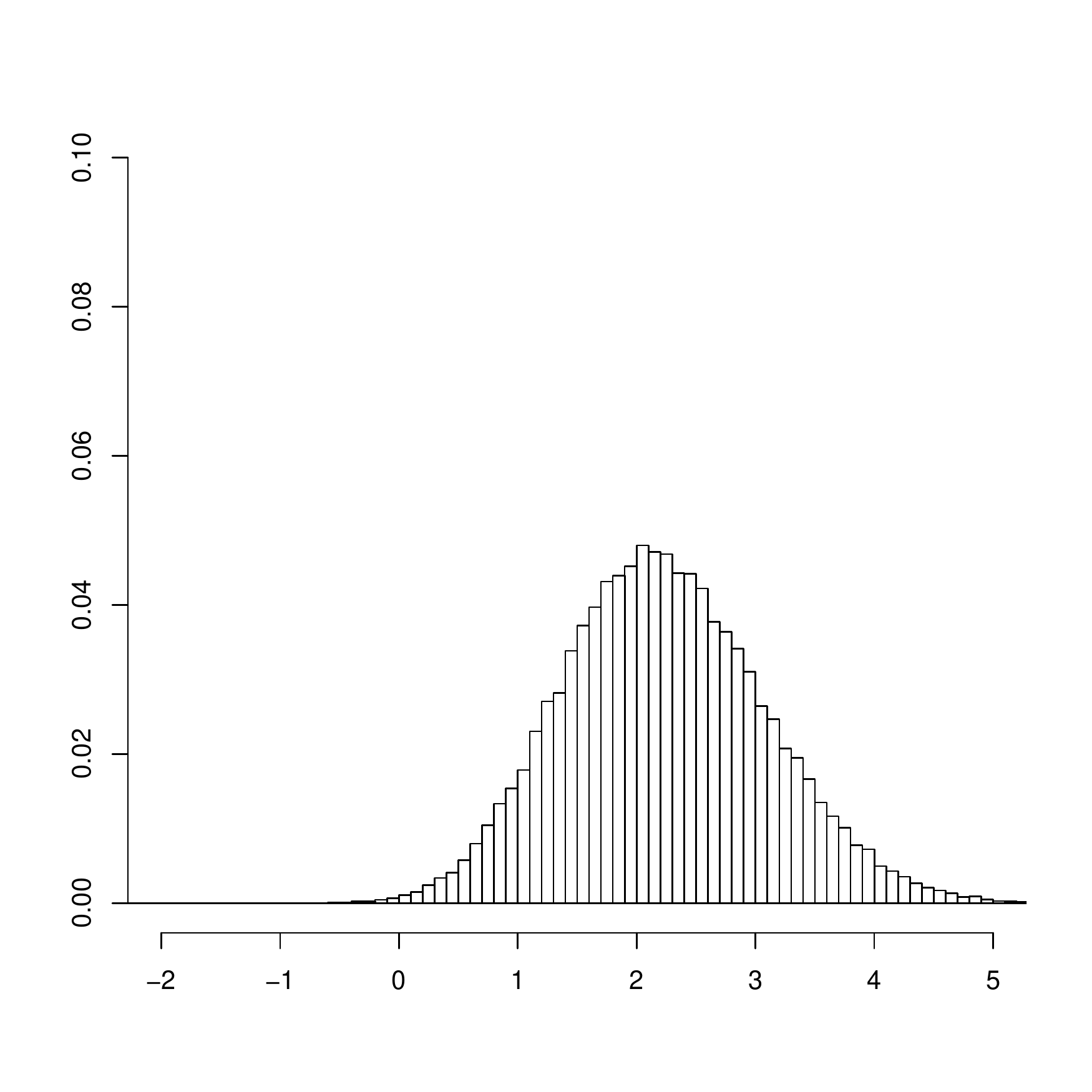}}
\subfigure[$\widehat\theta^c$, a mixture of $\widehat\theta_{(1)}^c$ and $\widehat\theta_{(2)}^c$]{\includegraphics[width = 2in,height=.18\textheight]{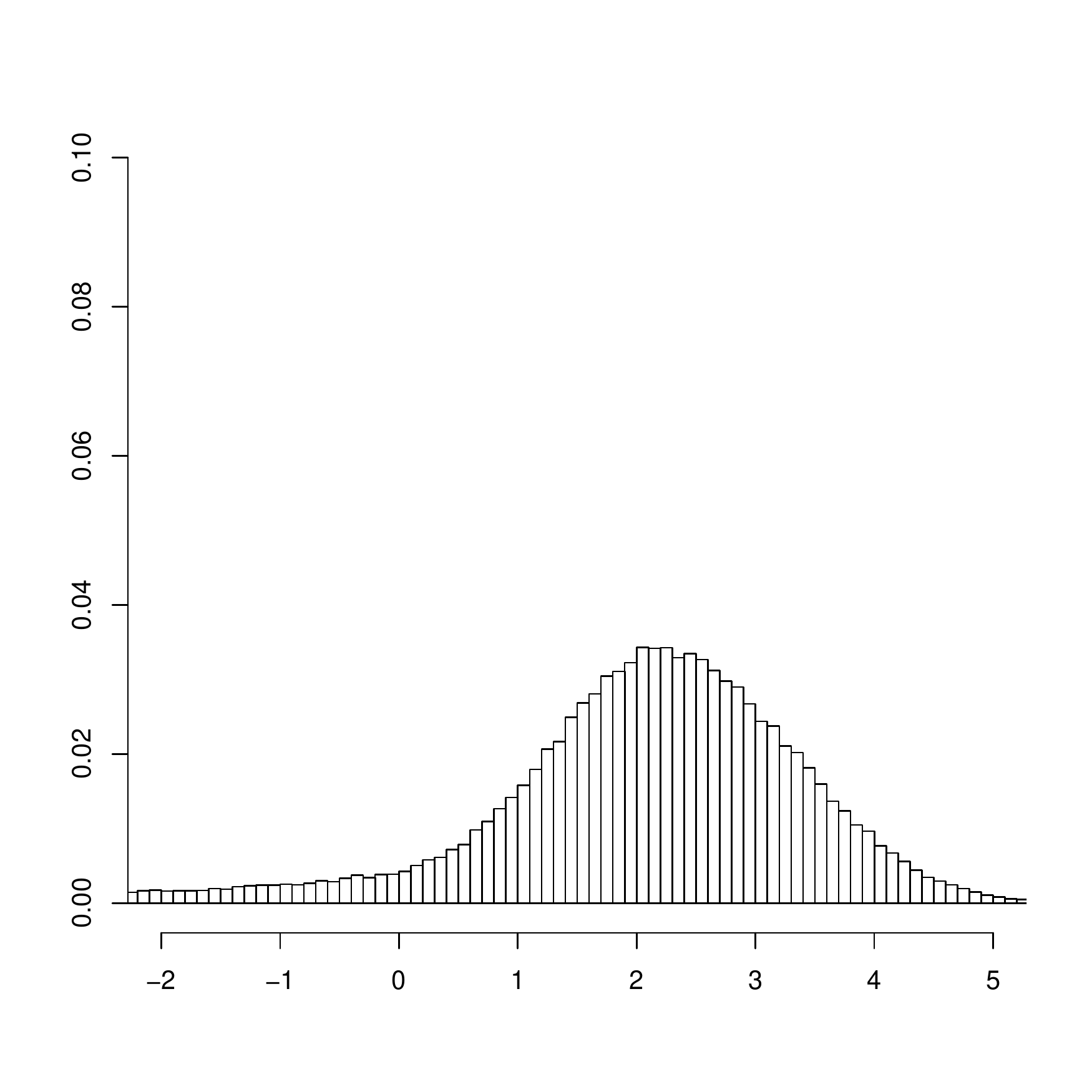}}
\caption{\label{Z_example} Histograms of MLEs for the Pocock example with critical value  $c_1=2.18$.}
\end{figure}

\begin{table}[ht!]
\centering
\begin{tabular}{c|c|c|c|c|c}
$\theta$ & Measure & $\widehat\theta_{(1)}$ & $\widehat\theta_{(1)}^c$ & 
                     $\widehat\theta_{(2)}$ & $\widehat\theta_{(2)}^c$ \\
\hline
0.0  & Bias & 0.2524 &-0.1865 & -0.0017 &  0.0025\\
0.0  & SD   & 0.0323 & 0.3136 &  0.0691 &  0.0741\\
0.0  & MSE  & 0.0647 & 0.1331 &  0.0048 &  0.0055\\
\hline
2.18 & Bias & 0.0796 &-0.1357 &  -0.0398 & 0.0056\\
2.18 & SD   & 0.0599 & 0.2972 &  0.0584 & 0.0852\\
2.18 & MSE  & 0.0099 & 0.1068 &  0.0050 & 0.0073\\
\end{tabular}
\caption{\label{table1} Monte-Carlo Simulation study ($100,000$ repetitions) 
for the two~stage Pocock design; $n_1=n_2=100$.}
\end{table}

\begin{figure}[ht!]
\centering
\subfigure[Study stops with stage~1. $Z_1=\sqrt{n_1} \bar X_1 \in \{2.3\text{(thin)},2.4,2.5,2.6\text{(thick)}\}$. ]{\includegraphics[width = 2.5in,height=2in]{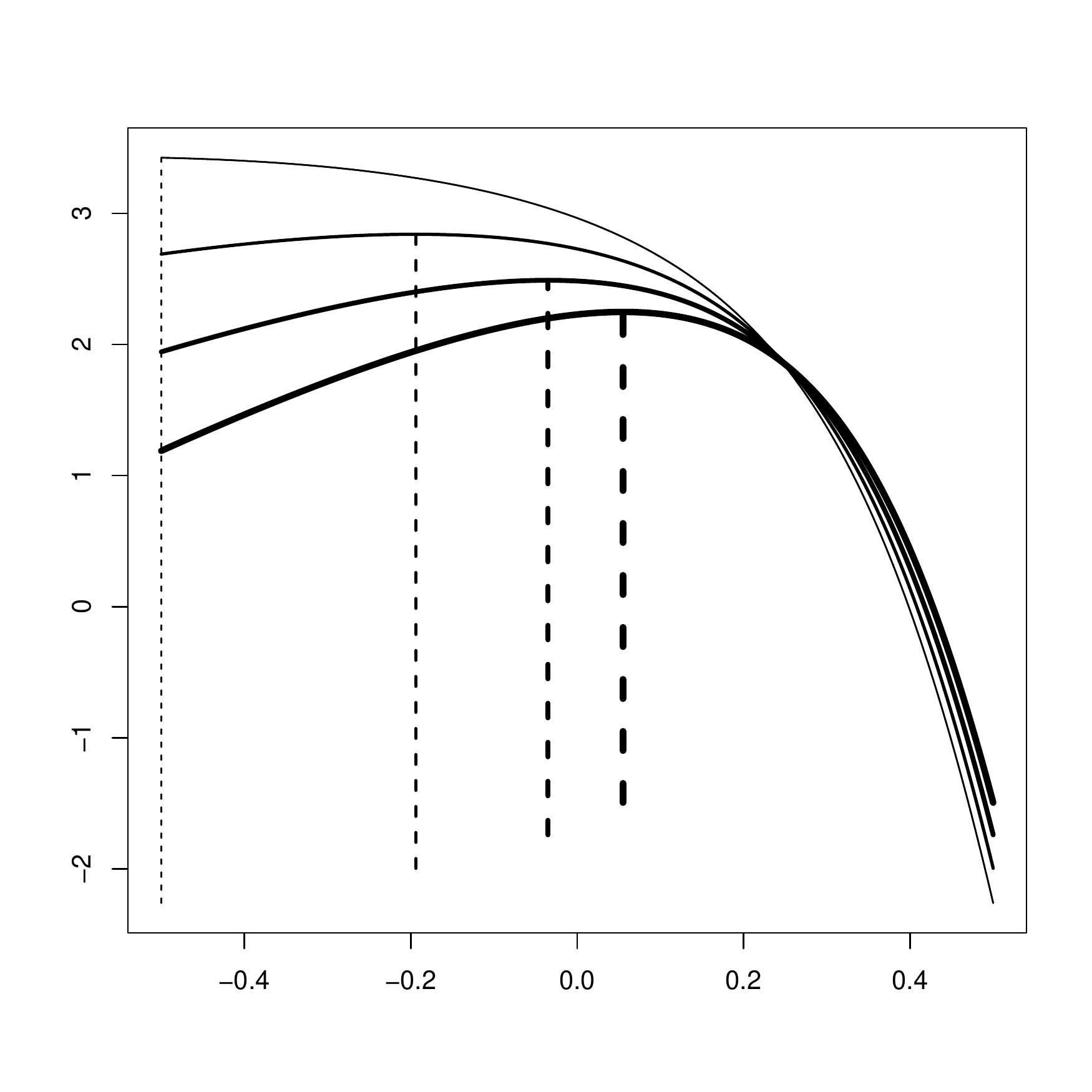}} \quad\quad
\subfigure[Study continues to stage~2. $Z_1=\sqrt{n_1} \bar X_1 \in \{2.0\text{(thin)},1.5,1.0,0.5\text{(thick)}\}$. ]{\includegraphics[width = 2.5in,height=2in]{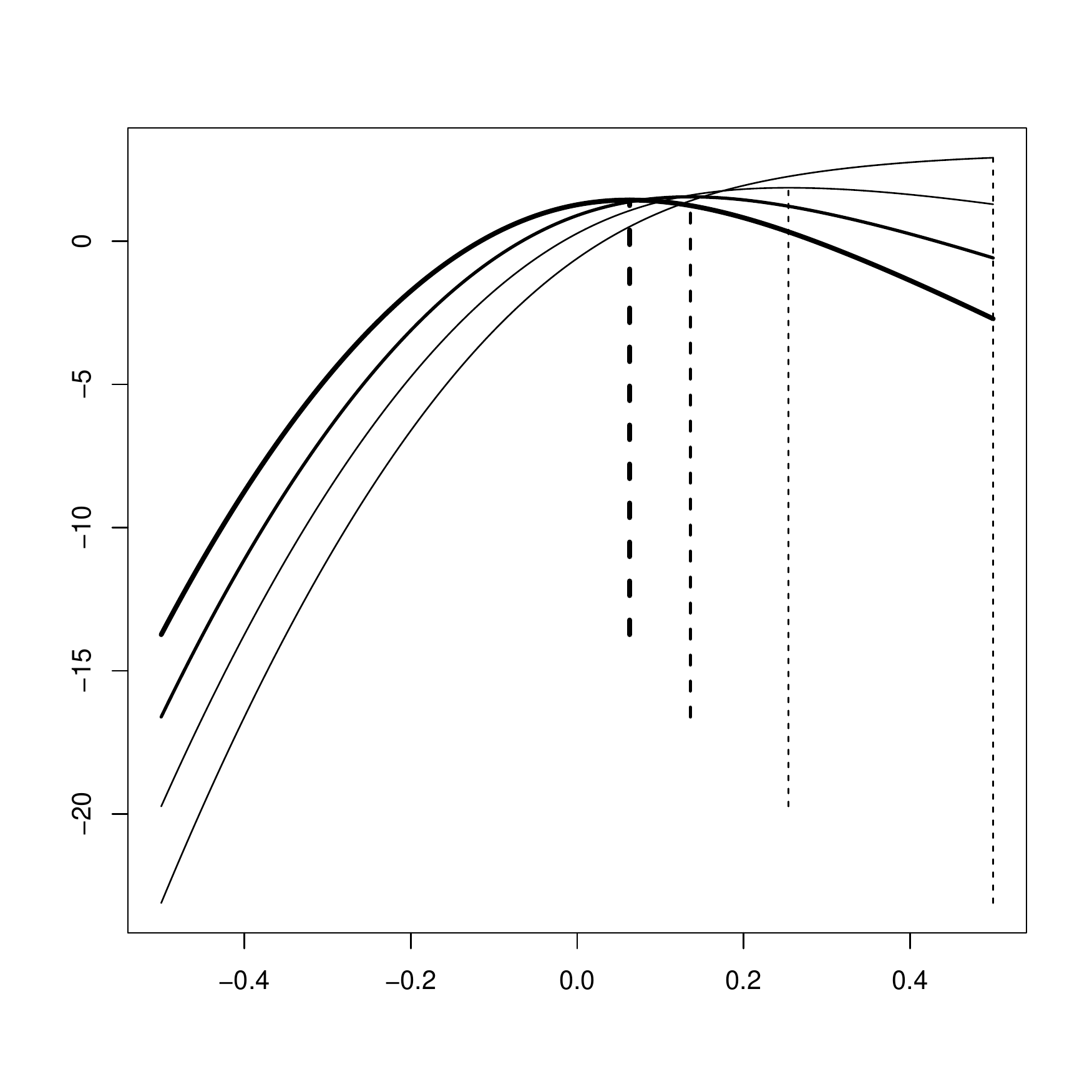}}\\
\caption{\label{likelihood}  Log-likelihood plots as a functions of $\theta$ given selected realizations  of $Z_1=\sqrt{n_1} \bar X_1$;
 $n_1=n_2=100$, $X \sim N(\theta,1)$; dotted lines trace maximums.  The critical value for stopping is $c_1= 2.18$. }
\end{figure}

\begin{figure}[ht!]
\centering
\subfigure[Information Measures from the Conditional Density: ${\cal{I}}_{(1)}^c$ and ${\cal{I}}_{(2)}^c$ (solid),
   ${\cal{I}}_{(1)}^{fix}$ and ${\cal{I}}_{(2)}^{fix}$ (dotted),
	 ${\cal{I}}_{(1)}$ and ${\cal{I}}_{(2)}$ (dashed); thick right is for stopping at stage 1 and a thick left for stage 2.]{\includegraphics[height=2in, width = 2.5in]{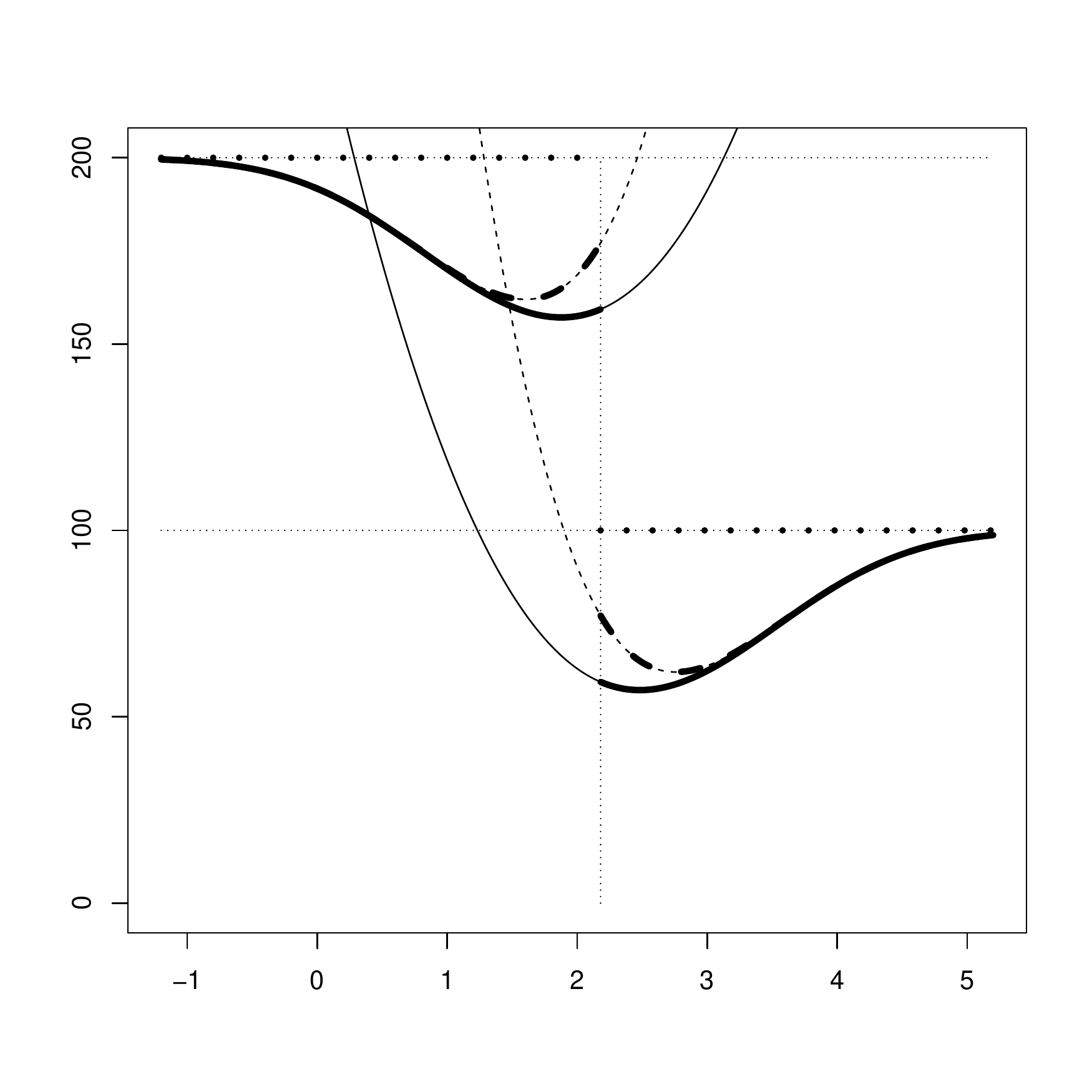}} \quad\quad
\subfigure[Information Measures from the Unconditional Density: ${\cal{I}}^c$ (solid), ${\cal{I}}^{fix}$ (dotted), ${\cal{I}}$  (dashed).]{\includegraphics[height=2in, width = 2.5in]{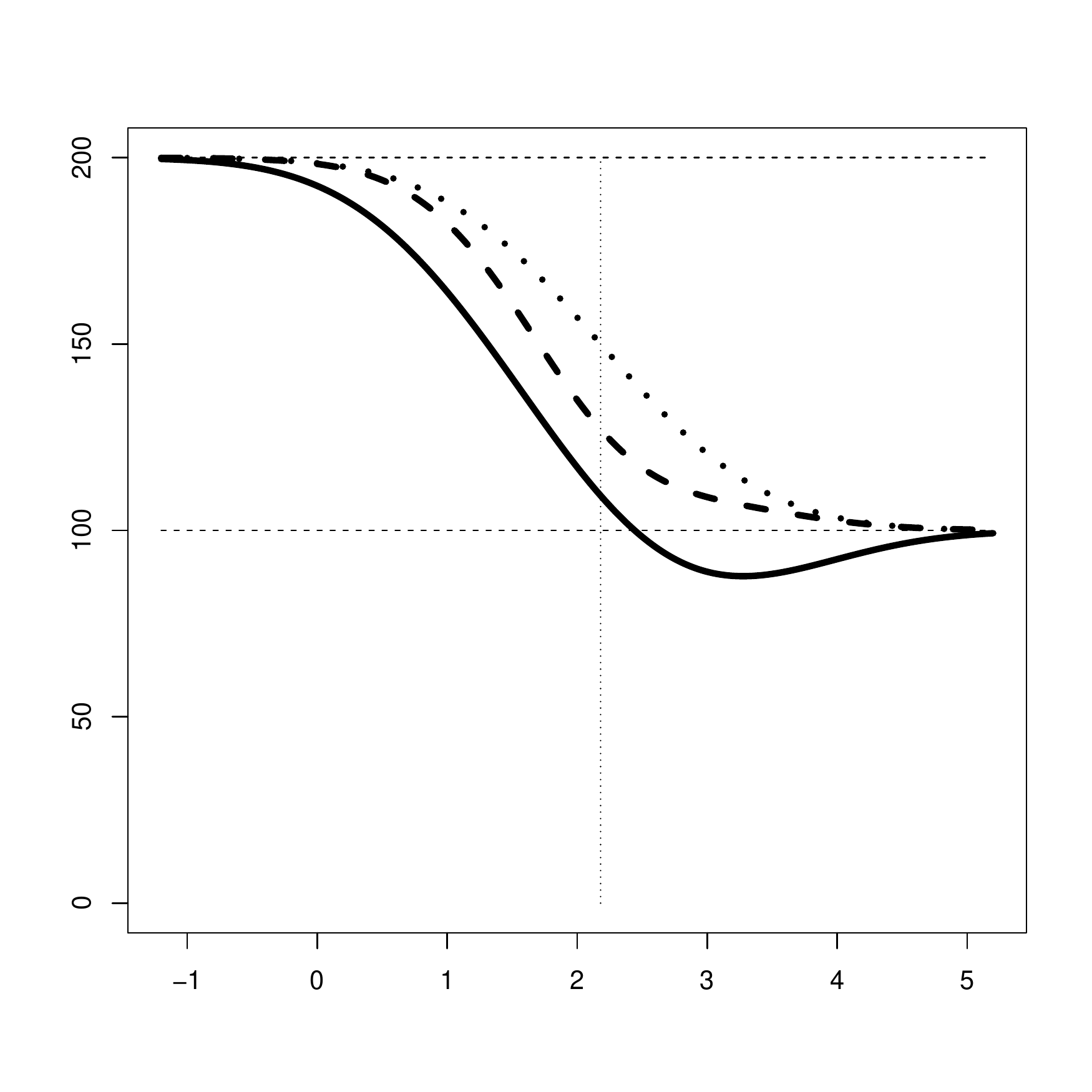}}\\
\caption{\label{FisherInf} Information Measures as a Function of $\theta$}
\end{figure}

Figure~\ref{Z_example} shows histograms of $\widehat\theta_{(1)}$, $\widehat\theta_{(1)}^c$, $\widehat\theta_{(2)}$, 
and $\widehat\theta_{(2)}^c$ estimated from $100,000$ Monte-Carlo samples assuming $\theta=2.18$.
Table~\ref{table1} reports Monte-Carlo biases, standard deviations (SD), and mean squared errors (MSE) under $\theta=0$ and under $\theta = 2.18$.
In this example, the study is stopped for efficacy at stage 1 if $Z_1>2.18$ and proceeds to stage 2 when $Z_1 \le 2.18$. If the study is stopped at stage~1,  the support for $\bar X_1$ starts at $2.18\sqrt{100}=0.218 $ and stretches to $+\infty$. If the study continues through stage~2, support for $\bar X_1$  ranges from $-\infty$ to $0.218$. However, it is still possible to obtain a conditional MLE outside the range of support, that is, above or below $0.218$ at  stage 1. 
Figure \ref{likelihood} shows that conditional likelihoods do not allow accurate estimation of $\theta$ when observed $Z_1$ is close to the critical value for deciding to stop, and the conditional MLEs start diverging to $\pm \infty$.

In GSDs, the ``information fraction'' is used to justify times or sample sizes of interim analyses. In two-stage designs, the information fraction is often estimated as ${\cal{I}}_{(1)}^{fix}/{\cal{I}}_{(2)}^{fix}$, which in  the Pocock example is equal to $1/2$ since ${\cal{I}}_{(1)}^{fix}=100$ and ${\cal{I}}_{(2)}^{fix} = 200$. This information fraction argument silently assumes that study stops at a pre-determined sample size. In Figure \ref{FisherInf}, ${\cal{I}}_{(1)}^{fix}$ and ${\cal{I}}_{(2)}^{fix}$ are plotted as a function of $\theta$ with dotted horizontal lines. Figure \ref{FisherInf} also graphs information in conditional and unconditional likelihoods.

\section{New Group Sequential Designs Relying on Mixture Distributions} \label{mixtureD}

Even though the normality does not hold when there is a possibility of early stopping, the critical values for controlling type~1 error are  correctly calculated. With the use of Monte-Carlo methods, data generated from stage-specific normal distributions with GSD-determined early stopping inherently generate draws from correct mixtures. Similarly, numeric integration [see \citet{jennison1999}] also correctly estimates type  error and power by the use of the recursive sub-density formula of \citet{Armitage1969}.

Currently, the design of many group sequential trials starts with choosing an $\alpha$-spending function. Then, ``Information fraction'' considerations are  used to define times or sample sizes of interim analyses that secure the overall power at a desired level. Statistical power to detect one or another treatment effect is only described. It is important to note that stage-specific sample sizes in such study designs are not driven by desired stage-specific power requirements, but fully determined by a (1) chosen spending functions, (2) ``information fraction'' for interim looks, and (3) a ``maximum information'' to reach a desired overall power.

In contrast, this manuscript proposes GSDs driven by \textit{a sequence of ordered alternative hypotheses} attainable with a predetermined statistical power while controlling type~1 error. 

\subsection{New GSDs Determined by  Ordered Alternatives}

Consider a study design in which stage-specific sample sizes $\{n_k\}$ and and critical values $\{c_k\}$ are determined by an $\alpha$-spending function and at least $1-\beta$ power of rejecting  $H_0:\theta = 0$ for each alternative hypothesis $H_{Ak}:\theta = \theta_k$, where $\theta_{1} > \theta_{2} > \cdots > \theta_K > 0$. Specifically,
 $\{n_k\}$ and $\{c_k\}$ are identified recursively to satisfy $\alpha_{(k)} \le \alpha$ and under $\theta=\theta_k$, $1-\beta_{(k)}(\theta_k) = 1-\beta$. 
The first stage critical value $c_1$  and sample size $n_1$ are determined by $\alpha_1$ and by $\beta =\beta_1$ which depends on the first stage alternative hypothesis $\theta_1$.  If the same alternative and sample size were used for stage 2, the power at stage~2 would be greater than $1-\beta$. A smaller $\theta_2<\theta_1$ and $\alpha_2$ are chosen to define a stage~2 sample size $n_2$ and the critical value $c_2$ that will keep the power at $1-\beta$.  Additional decreasing hypotheses and $\alpha_k$ are used to determine samples sizes and critical values for following stages. Thus, this design is ``flexible''. The total number of stages, $K$, does not need to be pre-determined.
However, for ease of illustration, henceforth  a fixed $K$ is considered.
\begin{table*}
	\centering
		\begin{tabular}{|c|c|}
		\hline
		Feature & Mathematical Definition \\
		\hline
		\textit{overall type $1$ error} & $\alpha = \sum_{k=1}^K \alpha_k \prod_{j=1}^{k-1}\left(1-\alpha_j\right)$\\
		\textit{stage-specific type 1 error} & $\alpha_k = \text{Pr}_{\theta}\left( Z_{(k)} > c_k \big| \cap_{j=1}^{k-1} \left\{ Z_{(j)}\le c_{j} \right\}, \theta = 0 \right)$ \\
		\textit{$\alpha$-spending function} & $\alpha_{(k)} = \sum_{j=1}^k\alpha_j \prod_{i=1}^{j-1}\left(1-\alpha_i\right)$\\
    \hline
		\textit{overall power} & $1-\beta(\theta) = \sum_{k=1}^K [1-\beta_k(\theta)] \prod_{j=1}^{k-1}\left[\beta_j(\theta)\right]$\\
		\textit{stage-specific power} & $  1-\beta_k(\theta) = \text{Pr}_{\theta}\left( Z_{(k)} > c_k \big| \cap_{j=1}^{k-1} \left\{ Z_{(j)}\le c_{j} \right\}, \theta \right)$\\
		\textit{cumulative power} & $1-\beta_{(k)}\left(\theta\right) = \sum_{j=1}^k[1-\beta_j(\theta)] \prod_{i=1}^{j-1}\left[\beta_i(\theta)\right]$\\
		\hline
		\textit{overall type 2 error} & $\beta(\theta)$	\\
		\textit{stage-specific type 2 error} & $\beta_k(\theta)$\\
		\textit{$\beta$-spending function} & $\beta_{(k)}\left(\theta\right)$\\
		\hline
				\textit{stage-specific rejection region} & $ {\cal{R}}_k = \left\{Z_{(k)} > c_k \right\} \cap \left\{\cap_{j=1}^{k-1} \left\{ Z_{(j)}\le c_{j} \right\} \right\}$ \\
				\textit{stage-specific rejection probability} & $  \text{Pr}_{\theta}\left({\cal{R}}_k \vert \theta \right)$\\
		\hline
		\end{tabular}
	\caption{Definitions of GSD operational characteristics ($k=1,\ldots,K$); $\prod_{i=1}^0[\cdot] = 1$.}
	\label{Definitions}
\end{table*}

The use of both $\alpha-$ and $\beta-$ spending functions was suggested by \citet{pampallona2001} for use with a single alternative hypothesis at predetermined samples sizes. But the use of a sequence of ordered alternative hypotheses with the same power to determine sample sizes is new. 
In this optimization problem, there are $K+1$ (marginal) restrictions (type 1 error $=\alpha$ under $H_0$ and power $=1-\beta$ under each alternative) and $2K$unknown quantities $\{n_1,\ldots,n_K,c_1,\ldots,c_K\}$. 
The $Z_{(k)}$ pivots do not have to be  normal random variables.  More details are given through the example that follows.
 
\subsection{One Simple Option}
It is common to select an $\alpha$-spending function that determines conditional type $1$ errors while keeping the overall type $1$ error equal to $\alpha$. 
For simplicity of exposition,  for all $k$, assume that $\alpha_k = \alpha_0$ leading to 
\begin{equation}\alpha = \sum_{k=1}^K \alpha_0 \prod_{j=1}^{k-1} \left(1-\alpha_0\right)= 
\alpha_0 \sum_{k=1}^K \left(1-\alpha_0\right)^{k-1} = \alpha_0\left(1-\left(1-\alpha_0\right)^K\right),\label{alphak2}
\end{equation} 
where the last equality follows from a property of geometric sequences.
Thus, for a fixed $K$, stage-specific critical values are defined via \eqref{alphak2}. For  a two-stage design with $\alpha = 0.05$,  $\alpha_0\approx 0.0257$; at $K=3$, $\alpha_0 \approx 0.0172$; at $K=4$, $\alpha_0 \approx 0.0144$.

The numerical example that follows shows how stage-specific power requirements can be determined from the overall power $1-\beta$ at $\theta = \theta_k$  requirement and sample sample sizes can be found that satisfy these requirements.  

\subsubsection{A numeric example at $K=3$.}

Consider $H_{A1}:\theta = 0.3$, $H_{A2}:\theta = 0.2$, and $H_{A3}:\theta = 0.1$. 
Stage-specific $n_k$ and $c_k$ $(k=1,2,3)$ are found from the following system of (nonlinear) equations:
\begin{eqnarray}\label{GSnew}
\text{Pr}_{\theta}\left(Z_{(1)}>c_1\big|\theta = 0\right) &=&  \alpha_0 \notag \\
\text{Pr}_{\theta}\left(Z_{(2)}>c_2\big|Z_{(1)}\le c_1,\theta = 0\right) &=&  \alpha_0\notag \\
\text{Pr}_{\theta}\left(Z_{(3)}>c_3\big|Z_{(2)}\le c_2, Z_{(1)}\le c_1,\theta = 0\right) &=&  \alpha_0\notag \\
\text{Pr}_{\theta}\left(Z_{(1)}>c_1\big|\theta = 0.3 \right) &=&  1-\beta \notag \\
\text{Pr}_{\theta}\left(Z_{(1)}>c_1\big|\theta = 0.2 \right) + \text{Pr}_{\theta}\left(\left\{Z_{(2)}>c_2\right\}\cap \left\{Z_{(1)}\le c_1\right\}\vert \theta = 0.2\right) &=&  1-\beta \notag \\
\sum_{k=1}^3 \text{Pr}_{\theta}\left(
\left\{Z_{(k)}>c_k \right\} \cap \left\{\cap_{1\le j \le k-1} \left\{ Z_{(j)}\le c_{j} \right\} \right\}\big| \theta = 0.1 \right) &=& 1-\beta
\end{eqnarray}
Recall that $Z_{(1)}$, $Z_{(2)}$ and $Z_{(3)}$ depend on the first $n_1$, $n_1+n_2$, and $n_1+n_2+n_3$ observations, respectively.
Resolving (\ref{GSnew}) numerically, under $X \sim N(\theta,1)$ and $\alpha_0 \approx 0.0172$, one finds $n_1=98$ and $c_1=2.12$ for stage 1, 
$n_2=98$ and $c_2 = 2.01$ for stage 2, and  $n_3=576$ and $c_3 = 2.02$ for stage 3. Using these stage-specific sample sizes and critical values, Table \ref{table_gs} reports operational characteristics of the new GSD based on a Monte-Carlo resampling with $100,000$ iterations.

\begin{table}[hbt]
\centering
\begin{tabular}{c|c|c|c|c|c}
$\theta$ & 
$\text{Pr}_{\theta}\left({\cal{R}}_1 \vert \theta \right)$ & 
$\text{Pr}_{\theta}\left({\cal{R}}_2 \vert \theta \right)$ & 
$\text{Pr}_{\theta}\left({\cal{R}}_3 \vert \theta \right)$ & $\text{E}[N\vert \theta]$ \\
\hline
0.0 & 0.0170 & 0.0336 & \textbf{0.0509} & $751$ \\
0.1 & 0.1287 & 0.3044 & \textbf{0.7983} & $584$ \\
0.2 & 0.4424 & \textbf{0.8016} & 0.9998 & $267$ \\
0.3 & \textbf{0.8018} & 0.9877 & 1.0000 & $125$ \\
\end{tabular}
\caption{\label{table_gs}  Stage-specific probabilities of rejecting $H_0$. $\text{E}[N\vert \theta]$ is the average sample size.}
\end{table}
As it will be shown in Section \ref{mostpower}, if each $Z_{(k)}$ is a continuous function of a monotone likelihood ratio, then the proposed test is the most powerful against any ordered alternatives $\theta_1 > \theta_2 > \cdots > \theta_K> 0$ given an $\alpha$-spending function. 

\subsection{Most Powerful Group Sequential Tests for Ordered Alternatives} \label{mostpower}
Let $X\sim f_X(\theta)$, where $f$ belongs to the exponential family. Without a possibility of early stopping, the likelihood for a realization $\textbf{\textit{x}}=\left(x_1,\ldots,x_n\right)$ of a random sample $\textbf{X}=\left(X_1,\ldots,X_n\right)$ is
\begin{align*}
{\cal{L}}\left(\theta|\textbf{\textit{x}}\right) = \prod_{i=1}^n f_X\left(\textbf{\textit{x}}\right) &= 
h\left(\textbf{\textit{x}}\right) g \left(T\left(\textbf{\textit{x}}\right) |\theta \right) = 
h\left(\textbf{\textit{x}}\right)  e^{\eta(\theta)T\left(\textbf{\textit{x}}\right) + A\left(\theta\right)}
\end{align*}
where all relevant information about $\theta$ is absorbed by a sufficient statistic $T\left(\textbf{\textit{x}}\right)$.  Assume the test statistic $Z$ is a one-to-one transformation of $T$.

If $LR(t):=g(t|\theta)/g(t|\theta_0)$ is a monotone likelihood ratio (MLR) in $t$, then the Karlin-Rubin theorem allows the construction of uniformly most powerful tests.
Using the notation in Section \ref{normality}, 
\begin{eqnarray}
{\cal{L}}^c\left(\theta|k,\textbf{\textit{x}}_{(k)}\right) 
&:=& \frac{I\left(D=k\right)}{\text{Pr}_{\theta}\left(D=k\right)}h\left(\textbf{\textit{x}}_{(k)}\right) 
e^{\eta(\theta)T\left(\textbf{\textit{x}}_{(k)}\right) + A\left(\theta\right)} \notag \\
&=& I\left(D=k\right) h\left(\textbf{\textit{x}}_{(k)}\right) 
e^{\eta(\theta)T\left(\textbf{\textit{x}}_{(k)}\right) + A\left(\theta\right) + \log \text{Pr}_{\theta}\left(D=k\right)}, \label{Lc}\\
{\cal{L}}\left(\theta|k,\textbf{\textit{x}}_{(k)}\right) 
&:=& I\left(D=k\right) h\left(\textbf{\textit{x}}_{(k)}\right) e^{\eta(\theta)T\left(\textbf{\textit{x}}_{(k)}\right) + A\left(\theta\right)},
\label{Lu}
\end{eqnarray} 
with associated likelihood ratios:
\begin{eqnarray}
LR^c(t) := \frac{{\cal{L}}^c\left(\theta|k,\textbf{\textit{x}}_{(k)}\right)}{{\cal{L}}^c\left(\theta_0|k,\textbf{\textit{x}}_{(k)}\right)} 
\quad\textrm{ and }\quad
LR(t) := \frac{{\cal{L}}\left(\theta|k,\textbf{\textit{x}}_{(k)}\right)}{{\cal{L}}\left(\theta_0|k,\textbf{\textit{x}}_{(k)}\right)}.
\notag
\end{eqnarray}
For every $D=k$, $LR^c(t) = LR(t)= \exp \left[\left(\eta(\theta)-\eta(\theta_0)\right) T_{(k)} + \left(A(\theta)-A(\theta_0)\right)\right]$ which means that \textit{the MLR property is preserved with early stopping}. 

\begin{theorem} \label{theorem1} For any fixed $(\alpha_1,\ldots,\alpha_K)$ and $\left(n_1,\ldots,n_K\right)$ $(1)$ $\{T_{(k)} > c_k\}$ is a UMP $\alpha_k$-level test and $(2)$ no test is more powerful than $\{T_{(D)} > c_D\}$.
\end{theorem}
\textbf{Proof.} 
$(1)$ As shown above,  conditional on $D=k$, $T_{(k)}$ is  sufficient  in exponential families and the MLR property is  preserved. By the Karlin-Rubin theorem, the test based on $T_{(k)}$, is uniformly most powerful.  
\\
$(2)$ 
Using Table \ref{Definitions} notation, 
$$1-\beta(\theta) 
= \sum_{k=1}^K \left(\prod_{j=1}^{k-1}\left[\beta_j(\theta)\right] \right) [1-\beta_k(\theta)]= 1 - \prod_{k=1}^{K}\beta_k(\theta).$$
At $K=1$, $\alpha_1$ and $n_1$ uniquely define $c_1$ and $\left\{T_{(1)} > c_{1}\right\}$ is a UPM by part 1 with a stage-specific power curve $1-\beta_1(\theta)$. 
At an arbitrary $K$, $\alpha_k$ and $n_k$ uniquely define $c_k$ and, by part 1 of the theorem, the stage-specific power $1-\beta_k(\theta)$, is the highest. Consequently, for any choice of $D$, the power of the test $\left\{T_{(D)} > c_{D}\right\}$ is $1-\beta(\theta) = 1-\prod_{k=1}^K\beta_k(\theta)$. This power is the highest, because the stage specific type 2 errors $\beta_k(\theta)$ are the lowest for each $k$ at any $\theta$. 
\hfill \textbf{Q.E.D.}\\[2pt]

Even if $T_k$ is normal, the distributions of $T_{(k)}$ are not normal,  but Armitage's recursive  sub-density formula can be used to evaluate the distribution of $T_{(k)}$. For example, at $K=2$, the sub-density of $T_{(2)}$ is $f^{sub}_{T_{(2)}}(t|\theta) = \int_{-\infty}^{c_1} f_{T_{(2)}|T_{(1)}} (t|t_1,\theta) f_{T_{(1)}}(t_1|\theta)  dt_1$ and its density is
$f_{T_{(2)}}(t|\theta) = f^{sub}_{T_{(2)}}(t|\theta)\left(\int_{-\infty}^{c_1} f_{T_{(1)}}(t_1|\theta)dt_1\right)^{-1}$. 
Recursively, the density conditional on reaching the $k^{th}$ interim analysis is
$$f_{T_{(k)}}(t|\theta) = \frac{\int_{-\infty}^{c_{k-1}} f_{T_{(k)}|T_{(k-1)}}(t|t_{k-1},\theta) f_{T_{(k-1)}}(t_{k-1}|\theta)  dt_{k-1}}{\int_{-\infty}^{c_{k-1}} f_{T_{(k-1)}}(t_{k-1}|\theta)dt_{k-1}}.$$ 

This section has shown the impact of the possibility of early stopping  on the finite sample distribution of test statistics. Section~\ref{LSP}  shows how interim decisions to stop or continue affect asymptotic properties.

\section{Large Sample Properties of Parameter Estimates}\label{LSP} 

Because unconditional MLEs ($\widehat\theta$) retain more information and have smaller MSEs than conditional MLEs ($\widehat\theta^c$), henceforth only unconditional MLEs are considered and they are now referred to simply as MLEs. To describe local asymptotic characteristics, the approach considered for sample size recalculation in \citet{Tarima2019} is adopted. Let $f(x|\theta,\eta)$ depend on a parameter of interest, $\theta$, and a nuisance parameter, $\eta$. The objective is to test a null $H_0:\theta=0$ versus local alternatives $H_{Ak}: \theta = h_{k}/\sqrt{n_1}$. 
The $k$th stage-specific estimates $(\widehat\theta_k,\widehat\eta_k)$ of $(\theta,\eta)$ and their statistical models are called 
\emph{regular} if, without the possibility of early stopping,
\begin{equation} \label{ctlholds}
\sqrt{n_k}\left(\begin{array}{c}
	\widehat\theta_k - \theta \\
	\widehat\eta_k - \eta
\end{array}
\right) \overset{d}{\to} N\left[
\left(\begin{array}{c}
	0\\
	0
\end{array}
\right),
\left(\begin{array}{cc}
	\sigma_{\theta\theta} & \sigma_{\theta\eta}\\
	\sigma_{\theta\eta} &  \sigma_{\eta\eta}
\end{array}
\right)
\right],
\end{equation}
where the limiting covariance matrix is positive definite with finite elements. Then,
\begin{equation} \label{proMLEdist}
\xi_k:= \sqrt{n_k}\left(\widehat\theta_k - \theta\right) \sigma_{\theta}^{-1} \overset{d}{\to} N\left(0, 1 \right), \qquad k=1,\ldots,K,
\end{equation}
where $\sigma_{\theta}^2 := \sigma_{\theta\theta}$. 
Assumption \eqref{ctlholds} was
described  in \citet{Tarima2019} to  encompass 
specific assumptions for
\begin{itemize} 
\item  independent, identically distributed observations by Cram\'er [e.g., for example, \citet{Ferguson1996}], 
\item  independent not identically distributed observations [e.g., \citet{philippou1973asymptotic}], 
\item  dependent observations  [e.g., \citet{crowder1976maximum}], 
\item and  densities whose support depends on parameters [e.g., \citet{Wang14}]. 
\end{itemize}
All these specific sets of assumptions include assumptions of the existence and consistency of the MLE. 

\subsection{Asymptotic Results under Local Alternatives for Two-Stage Group Sequential Designs}
With large sample sizes, due to  \eqref{ctlholds}, MLEs calculated on $n_1+n_2$ independent and identically distributed observations can be approximated using 
$$\tilde\theta_{(2)} = \frac{n_1}{n_1+n_2}\widehat\theta_1 + \frac{n_2}{n_1+n_2}\widehat\theta_2 + o_p\left(n_1^{-0.5}\right)$$
if data collection did not stop at stage $1$, otherwise use $\tilde\theta_{(1)} = \widehat\theta_1$. 
Then, with a random stopping index $D$, 
$$\tilde\theta_{(D)} = I(D=1)\tilde\theta_{(1)} + I(D=2)\tilde\theta_{(2)}.$$

As shown in Theorem 1 in \citet{Tarima2019}, the asymptotic properties of the standardized estimate
\begin{align}
V_{(D)} &:= \sqrt{n_{(D}}\left(\tilde\theta_{(D)} - \theta\right)/\sigma_{\theta} \notag \\
&= I(D=1)\sqrt{n_1}\left(\tilde\theta_{(1)} - \theta\right)/\sigma_{\theta} + 
I(D=2)\sqrt{n_1+n_2}\left(\tilde\theta_{(2)} - \theta\right)/\sigma_{\theta}\end{align} depend on the existence and distribution of the limiting random variable $r_{(D)}$ that is defined by
$$I(D=1) + I(D=2)\frac{n_1+n_2}{n_2} \overset{d}{\to} r_{(D)},$$
as $n_1\to\infty$. Then, for a sequence of local alternative hypotheses $\theta = h/\sqrt{n_1}$,
\begin{eqnarray}\text{Pr}_{\theta}\left(V_{(D)} < v \right) &\to& p_1 \Phi\left(\left. v\right\vert D=1\right) \notag \\
&+& (1-p_1) \int_{-\infty}^{c_1} \Phi\left(
\sqrt{r_{(D)}} v - \sqrt{r_{(D)}-1} y\right) \phi\left(y|D=2\right) dy,
\end{eqnarray} where $p_1 = \lim_{n_1\to\infty}{\text{Pr}_{\theta}}\left(D=1\right)$ is the limiting stage 1 stopping probability. Using previous terminology, under $\theta = 0$, $V_{(2)}=Z_{1}=Z_{(1)}$ if stopped at stage 1, and $V_{(2)}=Z_{(2)}$, if the study proceeds to the second stage.

\subsection{Example: The Two-Stage One-Sided Pocock Design}
If $n_1=n_2$, then $r_{(D)} = 1 + I(D=2)$ and $V_{(D)} = I(D=1)Z_1 + I(D=2)\left(Z_1 + Z_2\right)/\sqrt{2}$. Then, 
$$\text{Pr}_{\theta}\left(V_{(D)} < v \right) \to (\alpha/2) \Phi\left(\left. v\right\vert Z_1>c_1 \right)
+ (1-\alpha/2) \int_{-\infty}^{c_1} \Phi\left(
\sqrt{2} v - y\right) \frac{\phi\left(y\right)}{\text{Pr}\left(y \le c_1\right)}dy.
$$
Note if $Z_1\le c_1$, then $$\text{Pr}_{\theta}\left(\frac{Z_1+Z_2}{\sqrt{2}} < v \vert Z_1\le c_1\right) = 
\int_{-\infty}^{c_1} \Phi\left(
\sqrt{2} v - y\right) \frac{\phi\left(y\right)}{\text{Pr}\left(y \le c_1\right)} dy,$$
which is a continuous mixture of distributions.

\subsection{Asymptotic Results under Local Alternatives for $K$-Group Sequential Designs }
With large samples, the MLE given $D=k$ can be approximated recursively by
\begin{eqnarray}
\tilde\theta_{(k)} &\approx& \frac{n_{(k-1)}}{n_{(k)}}\tilde\theta_{(k-1)} + 
                 \frac{n_k}{n_{(k)}}\widehat\theta_k \label{MLE_k}
								\approx \sum_{j=1}^k \frac{n_j}{n_{(k)}} \widehat\theta_j,
\end{eqnarray}
where  $\tilde\theta_{(k-1)}$ is an MLE based on cumulative data from stages $1$ to $k-1$, and $\widehat\theta_{k}$ is the MLE based on stage $k$ data only.
In $K$-stage GSDs, the limiting random variable $r_{(D)}$  generalizes to 
\begin{equation}\label{tau_n}
\sum_{k=1}^K I(D=k)\frac{n_{(k)}}{n_{k}} \overset{d}{\to} \sum_{k=1}^K I(D=k)r_{(k)} := r_{(D)},
\end{equation} 
where $r_{(k)}$ is the asymptotic ratio of a cumulative and a stage-specific sample size. 
Equation (\ref{tau_n}) generalizes a $2$-dimensional definition of $r_{(D)}$. Thus, $r_{(D)}$ is a multinomial random variable with support on $r_{(k)}$, $k=1,\ldots,K$. 

Let $n_{(k)}/n_j \to r_{(k)j} \in (0,\infty)$, $j\le k$ with $r_{(k)k} = r_{(k)}$, then equations (\ref{MLE_k}) and (\ref{tau_n}) lead to the  standardized test statistic given  on $D=k$:
$$V_{(k)} = \sqrt{n_{(k)}} \sum_{j=1}^k \frac{n_j}{n_{(k)}} \left(\widehat\theta_j-\theta\right) =  
 \sum_{j=1}^k \sqrt{\frac{n_j}{n_{(k)}}} \sqrt{n_{j}} \left(\widehat\theta_j-\theta\right) \to \sum_{j=1}^k \frac{\xi_j}{\sqrt{r_{(k)j}}}.$$
Hence, the distribution of $V_{(D)}$ is
\begin{eqnarray}{\text{Pr}_{\theta}}\left(V_{(D)} < v \right) &\to& 
\sum_{k=1}^K {\text{Pr}_{\theta}}\left(r_{(D)}=r_{(k)}\right){\text{Pr}_{\theta}}\left(\sum_{j=1}^k \frac{\xi_j}{\sqrt{r_{(k)j}}}  < v \Big\vert r_{(D)}=r_{(k)} \right).
\end{eqnarray}

\section{Example} \label{example} 
To illustrate use of a sequence of ordered alternative hypotheses in designing a group sequential study, a dataset of $7,874$ persons available in Dr. Terry Therneau's ``Survival'' R package is reanalyzed. Serum free light chain (FLC), age, sex, creatinine, survival status and a follow-up period are available in this dataset. This smaller dataset is a subset of a larger dataset of $15,759$ persons where relationship between FLC and survival was analyzed using a proportional hazards model [\citet{dispenzieri2012}].

 A Cox proportional hazards regression model is used to analyze the effect of FLC on survival hazards controlling for age, sex and creatinine:
$$
\log h(t|FLC, A, M, C) = \log h_b(t) + \beta_1 FLC + \beta_2 A + \beta_2 M + \beta_3 C,
$$
where $h_b(t)$ is the baseline survival hazard; $A$ is ``Age''; $M$ is an indicator of male sex, and $C$ is a creatinine value.
When this model is applied to the whole $7,874$ patients, the regression coefficient of FLC ($\beta_1$) is estimated as $\widehat\beta_1 = 0.1368$ $(SE=0.0093)$ with $p<0.0001$.  Obviously, this large sample size was not needed to prove the significance of association. 

Consider a three stage sequential design simulated in Table \ref{table_gs} for detecting ordered standardized effect sizes $0.3$, $0.2$, and $0.1$ with 80\% power each while controlling overall overall type I error at 5\%. Per this design, $Z$-values need to be compared against the critical values of $2.12$ (stage 1, $n_1=98$), $2.01$ (stage 2, $n_2=98$), and $2.02$ (stage 3, $n=576$).
At stage 1, $\widehat\beta_1 = 0.1526$ $(SE=0.0520)$ with $p=0.0033$. The $Z$-statistic is $2.9379>c_1=2.12$. The study is stopped at stage 1 and the hypothesis of no association is rejected. 

To convert this example into a simulation study, this three-stage GSD is repeated $1,000$ times on the  randomly reshuffled FLC dataset. Overall the rejection rate was 74\% with the average sample size of 399. Study stopped at stage 1 in 35\% simulations and at stage 2 in 24\%.

\section{Summary}\label{Conclusion}

This paper focuses on GSDs with  stopping rules dependent on a parameter of interest. The information in conditional MLEs and unconditional MLEs is derived in Section \ref{normality}, and the  loss using conditional MLEs is found to be
${\cal{I}} - {\cal{I}}^c$  $= - \text{E}_D\left[\frac{\partial^2}{\partial\theta^2} \log  \text{Pr}_{\theta}\left(D=k\right)\right]\ge 0$. These findings show that ``information fraction'' should not be calculated using the amount of information in a sample with a fixed sample size $ {\cal{I}}^{fix}_{(k)} $. 
With normal random variables, ${\cal{I}}^{fix}_{(k)}/{\cal{I}}^{fix}_{(K)}= n_{(k)}/n_{(K)}$ . The correct ``information fraction'' is ${\cal{I}}_{(k)}/{\cal{I}}_{(K)}$, the use of which requires substantial computational effort. This difficulty is avoided with the use of the new GSD that is presented in Section \ref{mixtureD}. The new GSD ensures the desired power (for example, 80\%) if data came from any of the multiple ordered alternative hypotheses at a pre-determined overall type I error (for example, 5\%). Since clinical research community understands statistical power, this new approach will likely be easily adopted by clinicians.

The traditional assumption of multivariate normality does not hold in the presence of random stopping rules. Distributions of sufficient statistics $T_{(k)}$ that are normally distributed without possibility of early stopping  become non-normal in GSDs. Consequently, information measures and information fractions are changed as well. Despite their non-normality, nevertheless, $T_{(k)}$ continues to be sufficient and if monotone likelihood ratio holds without adaptation, it continues to hold with adaptation. This immediately leads to most powerful tests at a predefined $\alpha$-spending function. It is important to note that each choice of an $\alpha-$spending function ``cuts out'' a different subspace of the sample space with its own sub-$\sigma$-field and  probability measure on this ``cut-out'' measurable space. This is why Theorem \ref{theorem1} conditions on a pre-determined $\alpha$-spending function. 

Section \ref{LSP} reports local asymptotic properties of the MLEs presented in Sections \ref{normality} and \ref{mixtureD}. The distributions of GSD's MLEs are different from regular non-GSD MLEs. Their asymptotics, under  regularity conditions, is described by mixtures and truncated normal random variables. Finally, Section \ref{example} illustrates how this approach applies to GSD in Cox proportional hazards regression models.

Overall,  researchers need to be careful dealing with information fractions when designing sequential trials and  sequential designs powered for multiple ordered alternative hypotheses is recommended.

\bibliographystyle{Chicago}

\bibliography{bib}
\end{document}